\RequirePackage{fix-cm}
\documentclass[]{article}          
\usepackage{etex}

\usepackage[utf8]{inputenc}		
\usepackage[english]{babel}		
\usepackage[T1]{fontenc}		

\usepackage{geometry}	
\usepackage[left,modulo]{lineno}

\usepackage[fleqn]{empheq}	       
\usepackage{mathptmx}                
\numberwithin{equation}{section}
\usepackage{amssymb}                 
\usepackage[fleqn]{cases}             
\usepackage{mathrsfs}                 
\usepackage{bm}                          
\allowdisplaybreaks

\DeclareSymbolFont{calletters}{OMS}{cmsy}{m}{n}
\DeclareSymbolFontAlphabet{\mathcal}{calletters}

\usepackage[nointegrals]{wasysym}
\usepackage{stmaryrd}
\usepackage{esint}			
\usepackage[misc]{ifsym}
\usepackage{extarrows}

\usepackage[margin=1cm, font={small,it}]{caption}
\usepackage[]{graphicx}		
\usepackage{grffile}
\usepackage{float}
\usepackage[section]{placeins}
\usepackage{subfig}
\usepackage{wrapfig}		
\usepackage{epstopdf}		
\usepackage{xcolor}

\usepackage{enumitem}

\usepackage{array}			
\usepackage{dcolumn}    
\usepackage{booktabs}
\usepackage{tabu}
\usepackage{tabularx}
\usepackage{multirow}
\usepackage{diagbox}

\usepackage{tikz-cd}
\usetikzlibrary{shapes,arrows}
\usepackage{extarrows}

\usepackage{siunitx}

\usepackage[authoryear,round]{natbib}

\usepackage{appendix}	

\usepackage[draft]{pdfcomment}

\usepackage{verbatimbox}
\usepackage{xspace}

\usepackage{authblk}

\newcolumntype{.}{D{.}{.}{-1}}
\newcolumntype{R}[1]{>{\raggedright\arraybackslash$\displaystyle}p{#1}<{$}}
\newcolumntype{L}[1]{>{\raggedleft\arraybackslash$\displaystyle}p{#1}<{$}}
\newcolumntype{K}[1]{>{\centering\arraybackslash$\displaystyle}p{#1}<{$}} 
\newcolumntype{C}{>{\centering\arraybackslash$\displaystyle}c<{$}}

\newcolumntype{G}[1]{>{\raggedright\arraybackslash }b{#1}}

\newcommand{\mcpsc}{\si{\square\metre\per\square\second}}

\newcommand{\matr}[1]{\mathbf{#1}}

\DeclarePairedDelimiterX{\fracc}[2]{\lbrack}{\rbrack}{\dfrac{#1}{#2}}
\DeclarePairedDelimiterX{\fracp}[2]{\lparen}{\rparen}{\dfrac{#1}{#2}}
\DeclarePairedDelimiterX{\fracabs}[2]{\lvert}{\rvert}{\dfrac{#1}{#2}}

\newcommand{\grdist}{\delta{g}}
\newcommand{\Xref}[1]{\overline{#1}}
\newcommand{\Xest}[1]{\widetilde{#1}}

\begin{document}

\title{Determination of a high spatial resolution
	geopotential model using \textcolor{black}{atomic clock comparisons}}


\author[1,2]{G.~Lion\thanks{Guillaume.Lion@obspm.fr}}
\author[2]{I. Panet}
\author[1]{P.~Wolf}
\author[1,3]{C.~Guerlin}
\author[1]{S.~Bize}
\author[1]{P.~Delva}
\affil[1]{LNE-SYRTE, 
	Observatoire de Paris, 
	PSL Research University, 
	CNRS, 
	Sorbonne Universités,
	UPMC Univ. Paris~06,
	61 avenue de l’Observatoire, F-75014 Paris, France}
\affil[2]{LASTIG LAREG, IGN, ENSG, Univ Paris Diderot, Sorbonne Paris Cité,
	35 rue H\'el\`ene Brion, 75013 Paris, France}

\affil[3]{Laboratoire Kastler Brossel, ENS-PSL Research University, CNRS,
	UPMC-Sorbonne Universit\'es, Coll\`ege de France, 24 rue Lhomond, 75005 Paris, France}

\renewcommand\Authands{ and }

\date{}

\maketitle


\begin{abstract}
Recent technological advances in optical atomic clocks are opening new perspectives for the direct determination of geopotential differences between any two points at a centimeter-level accuracy in geoid height. However, so far detailed quantitative estimates of the possible improvement in geoid determination when adding such clock measurements to existing data are lacking. We present a first step in that direction with the aim and hope of triggering further work and efforts in this emerging field of chronometric geodesy and geophysics.
We specifically focus on evaluating the contribution of this new kind of direct measurements in determining the geopotential at high spatial resolution ($\approx 10$~\si{\km}). We studied two test areas, both located in France and corresponding to a middle (Massif Central) and high (Alps) mountainous terrain.
These regions are interesting because the gravitational field strength varies greatly from place to place at high spatial resolution due to the complex topography.
Our method consists in first generating a synthetic high-resolution geopotential map, then drawing synthetic measurement data (gravimetry and clock data) from it, and finally reconstructing the geopotential map from that data using least squares collocation. The quality of the reconstructed map is then assessed by comparing it to the original one used to generate the data. We show that adding only a few clock data points (less than 1~\% of the gravimetry data) reduces the bias significantly and improves the standard deviation by a factor~3. The effect of the data coverage and data quality on the results is investigated, and the trade-off between the measurement noise level and the number of data points is discussed.

\end{abstract}

\smallskip
\noindent \textbf{Keywords.} 
{Chronometric geodesy; High spatial resolution; Geopotential; Gravity field; Atomic clock; Least-squares collocation (LSC); Stationary covariance function}

\section{Introduction}
\label{intro}
%

Chronometry is the science of the measurement of time. As the time flow of clocks depends on the surrounding gravity field through the relativistic gravitational redshift predicted by Einstein \citep{Landau_1975aa}, chronometric geodesy considers the use of clocks to directly determine Earth's gravitational potential differences. Instead of using state-of-the-art Earth's gravitational field models to predict frequency shifts between distant clocks~(\cite{Pavlis_2003aa}, ITOC project\footnote{{http://projects.npl.co.uk/}}), the principle is to reverse the problem and ask ourselves whether the \textcolor{black}{comparison} of frequency shifts between distant clocks can improve our knowledge of Earth's gravity and geoid \citep{Bjerhammar_1985aa, Mai_2013aa, Petit_2014aa, Shen_2016aa, Kopeikin_2016aa}. For example, two clocks with an accuracy of~$10^{-18}$ in terms of relative frequency shift would detect a 1-cm geoid height variation between them, corresponding to a geopotential variation~$\Delta W$ of about~$0.1$~\mcpsc \citep[for more details, see e.g.][]{Delva_2013aa,Mai_2013aa,Petit_2014aa}.

%
%
Until recently, the performances of optical clocks had not been sufficient \textcolor{black}{to make applications in practice} for the determination of Earth's gravity potential. \textcolor{black}{However, ongoing quick developments of optical clocks are opening these possibilities.}
In 2010, \citet{Chou_2010aa} demonstrated the ability of the new generation of atomic clocks, based on optical transitions, to sense geoid height differences with a 30-cm level of accuracy. To date, the best of these instruments reach a stability of~$1.6 \times 10^{- 18}$ (NIST, RIKEN + Univ. Tokyo, \citet{Hinkley_2013aa}) after 7 hours of integration time. More recently, an accuracy of~$2.1 \times 10^{- 18}$ (JILA, \citet{Nicholson_2015aa}) has been obtained, equivalent to geopotential differences of 0.2~\mcpsc, or 2~cm on the geoid. \textcolor{black}{Recently, \citet{Takano_2016aa} demonstrated the feasibility of cm-level chronometric geodesy. By connecting clocks separated by~15~km with a long telecom fiber, they found that the height difference between the distant clocks determined by the chronometric leveling \citep[see][]{Vermeer_1983aa} was in agreement with the classical leveling measurement within the clocks’ uncertainty of 5~cm.}
\textcolor{black}{Other related work using optical fiber or coaxial cable time-frequency transfer can be found in \citep{Shen_2013aa, Shen_2015aa}}.

Such results \textcolor{black}{stress} the question of what can we learn about Earth's gravity and mass sources using clocks, that we cannot easily derive from existing gravimetric data. Recent studies address this question; for example, \citet{Bondarescu_2012aa} discuss the value and future applicability of chronometric geodesy for  direct geoid mapping on continents and joint gravity potential surveying to determine subsurface density anomalies. They find that a geoid perturbation caused by a 1.5~km radius sphere with 20 per cent density anomaly buried at 2~km depth in the Earth’s crust is already detectable by atomic clocks with present-day accuracy. They also investigate other applications, for earthquake prediction and volcanic eruptions \citep{Bondarescu_2015aa}, or to monitor vertical surface motion changes due to magmatic, post-seismic, or tidal deformations \citep{Bondarescu_2015ab,Bondarescu_2015ac}. 

Here we will consider the "static" or "long-term" component of Earth's gravity. Our knowledge of Earth's gravitational field is usually expressed through geopotential grids and models that integrate all available observations, globally or over an area of interest. These models are, however, not based on direct observations with the potential itself, which has to be reconstructed or extrapolated by integrating measurements of its derivatives. \textcolor{black}{Yet, this quantity is needed in itself, like using a high-resolution geoid as a reference for height on land and dynamic topography over the oceans \citep{Rummel_1988aa, Rummel_2002aa, Sanso_2002aa, Zhang_2008aa, Rummel_2012aa, Sanso_2013aa, Marti_2015aa}.}
	

The potential is reconstructed with a centimetric accuracy at resolutions of the order of 100 km from GRACE and GOCE satellite data \citep{Pail_2011aa,Bruinsma_2014aa}, and integrated from near-surface gravimetry for the shorter spatial scales. As a result, the standard deviation (rms) of differences between geoid heights obtained from a global high-resolution model as EGM2008, and from a combination of GPS/leveling data, reaches up to 10~cm in areas well covered in surface data \citep{Gruber_2009aa}. 
%
The uneven distribution of surface gravity data, especially in transitional zones (coasts, borders between different countries) and with important gaps in areas difficult to access, indeed limits the accuracy of the reconstruction when aiming at a centimetric level of precision. This is an important issue, as large gravity and geoid variations over a range of spatial scales are found in mountainous regions, and because a high accuracy on altitudes determination is crucial in coastal zones. Airborne gravity surveys are thus realized in such regions \citep{Johnson_2009aa,Douch_2015aa}; local clock-based geopotential determination could be another way to overcome these limitations.

In this context, here, we investigate to what extent clocks could contribute to fill the gap between the satellite and near-surface gravity spectral and spatial coverages in order to improve our knowledge of the geopotential and gravity field at all wavelengths. By nature, potential data are smoother and more sensitive to mass sources at \textcolor{black}{large scales} than gravity data, which are strongly influenced by local effects. Thus, they could naturally complement existing networks in sparsely covered places and even also contribute to point out possible systematic patterns of errors in the less recent gravity datasets. We address the question through test case examples of high-resolution geopotential reconstructions in areas with different characteristics, leading to different variabilities of the gravity field. We consider the Massif Central in France, marked by smooth, moderate altitude mountains and volcanic plateaus, and an Alps-Mediterranean zone, comprising high reliefs and a land/sea transition.

\textcolor{black}{Throughout this work, we will treat clock measurements as direct determinations of the disturbing potential $T$ (see below and Section 3 for details). We implicitly assume that the actual measurements are the potential differences between the clock location and some reference clock(s) within the area of interest. These measurements are obtained by comparing the two clocks over distances of up to a few 100 km. Currently two methods are available for such comparisons, fiber links \citep{Lisdat_2016aa} and free space optical links \citep{Deschenes_2016aa}. The free space optical links are most promising for the applications considered here, but are presently still limited to short (few km) distances. However, projects for extending these methods based on airborne or satellite relays are on the way, but still require some effort in technology development.}

The paper is organized as follows. In Section~\ref{sec:2}, we briefly summarize the method schematically. In Section~\ref{sec:3}, we describe the regions of interest and the construction of the high-resolution synthetic datasets used in our tests. In Section~\ref{sec:4}, we present the methodology to assess the contribution of new clock data in the potential recovery, in addition to ground gravity measurements. Numerical results are shown in Section~\ref{sec:5}. We finally discuss in Section~\ref{sec:6} the influence of different parameters like the data noise level and coverage.

\section{Method}
\label{sec:2}
%

\begin{figure}[!htbp]
	\centering
	\resizebox{0.6\linewidth}{!}{%
	
	\begin{tikzpicture}
	\tikzstyle{quadri}=[rectangle,draw,text=black,text width=3cm,text centered]
	\tikzstyle{es}=[rectangle,draw,very thick,rounded corners=4pt,fill=gray!20,text width=3cm,text centered]
	\tikzstyle{ell}=[ellipse,draw,very thick,rounded corners=4pt,text width=1.5cm,text centered]
	\tikzstyle{estun}=[->,>=latex,thick,dashed]
	\tikzstyle{suite}=[->,>=latex,thick,rounded corners=4pt]
	\node[quadri] (A) at (-4,2) {\textbf{Step 1}: Build synthetic field model};
	\node[quadri] (B) at (-4,0) {\textbf{Step 2}: Select data distribution and add noise};
	\node[quadri] (C) at (-4,-2) {\textbf{Step 3}: Make an assumption on the a priori gravity field and estimate a potential model};
	\node[es] (D) at (0,2) {\textbf{Reference model}~$\Xref{\grdist}$ and~$\Xref{T}$};
	\node[es] (E) at (0,0) {\textbf{Synthetic data}~$\grdist$ and~$T$};
	\node[es] (F) at (0,-2) {\textbf{Estimated model}~$\Xest{T}$};
	\node[ell] (G) at (4,0) {Compute residuals~$\delta = \Xest{T}-\Xref{T}$ };
	\draw[suite] (A) -- (D);
	\draw[suite] (D) -- (E);
	\draw[estun] (G) |- (D);
	\draw[suite] (B) -- (E);
	\draw[suite] (E) -- (F);
	\draw[suite] (C) -- (F);
	\draw[estun] (G) |- (F);
	\end{tikzpicture}
	}
    \captionsetup{width=0.6\linewidth}  
	\caption{Scheme of the numerical approach used to evaluate the contribution of atomic clocks to determine the geopotential.}
	\label{fig:methodo}
\end{figure}
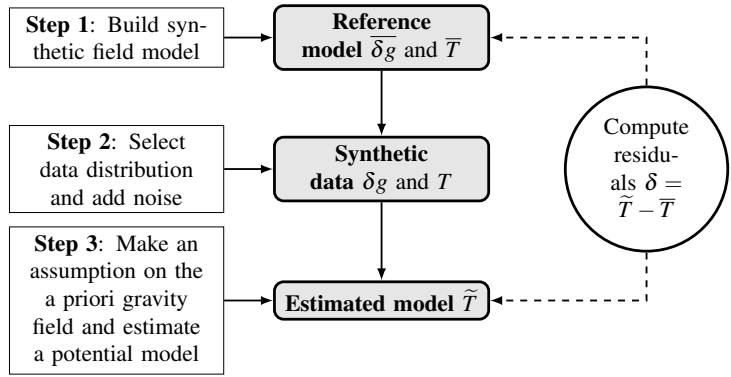

The rapid progress of optical clocks performances opens new perspectives for their use in geodesy and geophysics. While they were until recently built only as stationary laboratory devices, several transportable optical clocks are currently under construction or test (see e.g. \citet{Bongs_2015aa,Origlia_2016aa, Vogt_2016aa}.  The technological step toward state-of-the art transportable optical clocks is likely to take place within the next decade. In parallel, in order to assess the capabilities of this upcoming technology, we chose an approach based on numerical simulation in order to investigate whether atomic clocks can improve the determination of the geopotential. \textcolor{black}{Based on the consideration that ground optical clocks are more sensitive to the longer wavelengths of the gravitational field around them than gravity data}, our method is adapted to the determination of the geopotential at regional scales. In figure~\ref{fig:methodo} a scheme of the method used in this paper is shown:%
\begin{enumerate}
	\item In the first step, we generate a high spatial resolution grid of the gravity disturbance~$\Xref{\grdist}$ and the disturbing potential~$\Xref{T}$, considered as our reference solutions. This is done using a state of the art geopotential model (EIGEN-6C4), and by removing low and high frequencies. It is described in details in Section~\ref{sec:3};
	\item In the second step, we generate synthetic measurements~$\grdist$ and~$T$ from a realistic spatial distribution, then we add generated random noise representative of the measurement noise. This is described in details in Section~ \ref{sec:4};
	\item In a third step, we estimate the disturbing potential~$\Xest{T}$ from the synthetic measurements~$\grdist$ and/or~$T$ on a regular grid thanks to Least-Square Collocation (LSC) method. 
	\textcolor{black}{Interpolating spatial data} is realized by making an assumption on the a priori gravity field regularity on the target area, as described in Section~\ref{sec:5}.
	\textcolor{black}{This prior is expressed by the covariance function of the gravity potential and its derivatives. It allows to predict the disturbing potential on the output grid from the observations using the \textcolor{black}{signal} correlations between the data points, and with the estimated potential.}

	\item Finally, we evaluate the potential recovery quality for different data distribution sets, noise levels, and types of data, by comparing the statistics of the residuals~$\delta$ between the estimated values~$\Xest{T}$ and the reference model~$\Xref{T}$.
	
\end{enumerate}

\textcolor{black}{Let us underline that in this work, we use synthetic potential data while a network of clocks would give access to potential differences between the clocks. We indeed assume that the clocks-based potential differences have been connected to one or a few reference points, without introducing additional biases larger than the assumed clock uncertainties. Note that these reference points are absolute potential points determined by other methods (GNSS/geoid for example).}

\textcolor{black}{In this differential method, significant} residuals~$\delta$ (higher than the machine precision) can have several origins, depending on the parameters of the simulation that can be \textcolor{black}{varied}:%
\begin{enumerate} 
	\item The modeled instrumental noise added to the reference model at step 2. This noise can be changed in order to determine, for instance, whether it is better to reduce gravimetry noise by one order of magnitude, rather than using clock measurements;
	\item The data distribution chosen in step 2. This is useful to check for instance the effect of the number of clock measurements on the residuals or to find an optimal coverage for the clock measurements;
	\item The potential estimation error, due to the intrinsic imperfection of the \textcolor{black}{covariance model} chosen for the geopotential. In our case, this is due to the low-frequency content of the covariance function chosen for the Least-Square Collocation method (see Section~\ref{sec:5}).
\end{enumerate}
All these sources of errors are somewhat entangled with one another, such that a careful analysis must be done when varying the parameters of the simulation. This is discussed in details in Section~\ref{sec:6}.

\section{Regions of interest and synthetic gravity field reference models}
\label{sec:3}

\subsection{Gravity data and distribution}
\label{ssec:grav_data_distrib}

\begin{figure}[!htbp]
	\captionsetup{width=0.4\linewidth}  
	\centering
	\null\hfill
	\subfloat[Topography.]{%
		\includegraphics[clip=true, trim = 0 0 0 0, width=0.33\linewidth]{%
			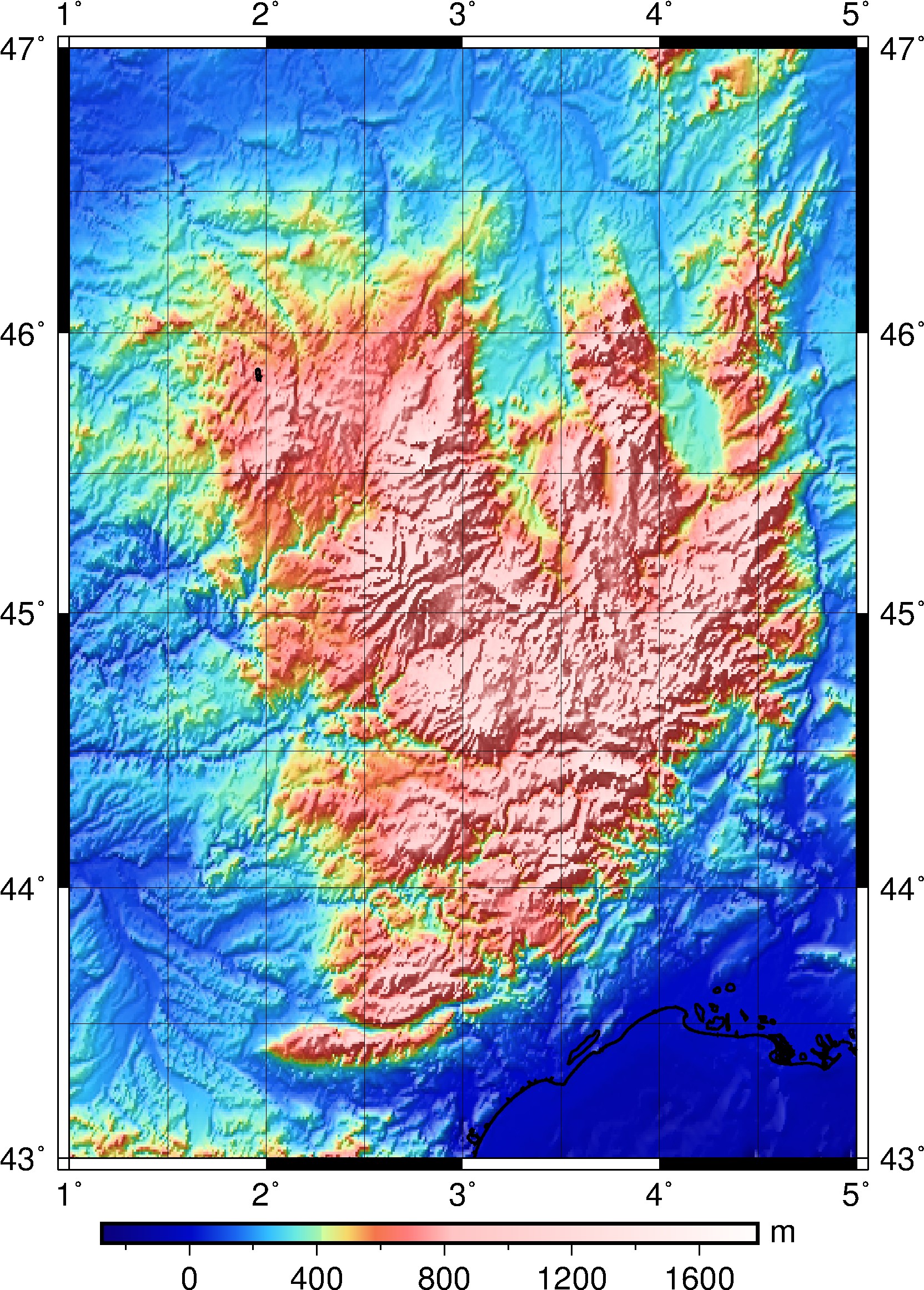}
		\label{fig:Auv_topo}}
	\hfill
	\subfloat[Terrestrial and marine free-air gravity anomalies.]{%
		\raisebox{0\height}{\includegraphics[clip=true, trim = 0 0 0 0, width=0.33\linewidth]{%
				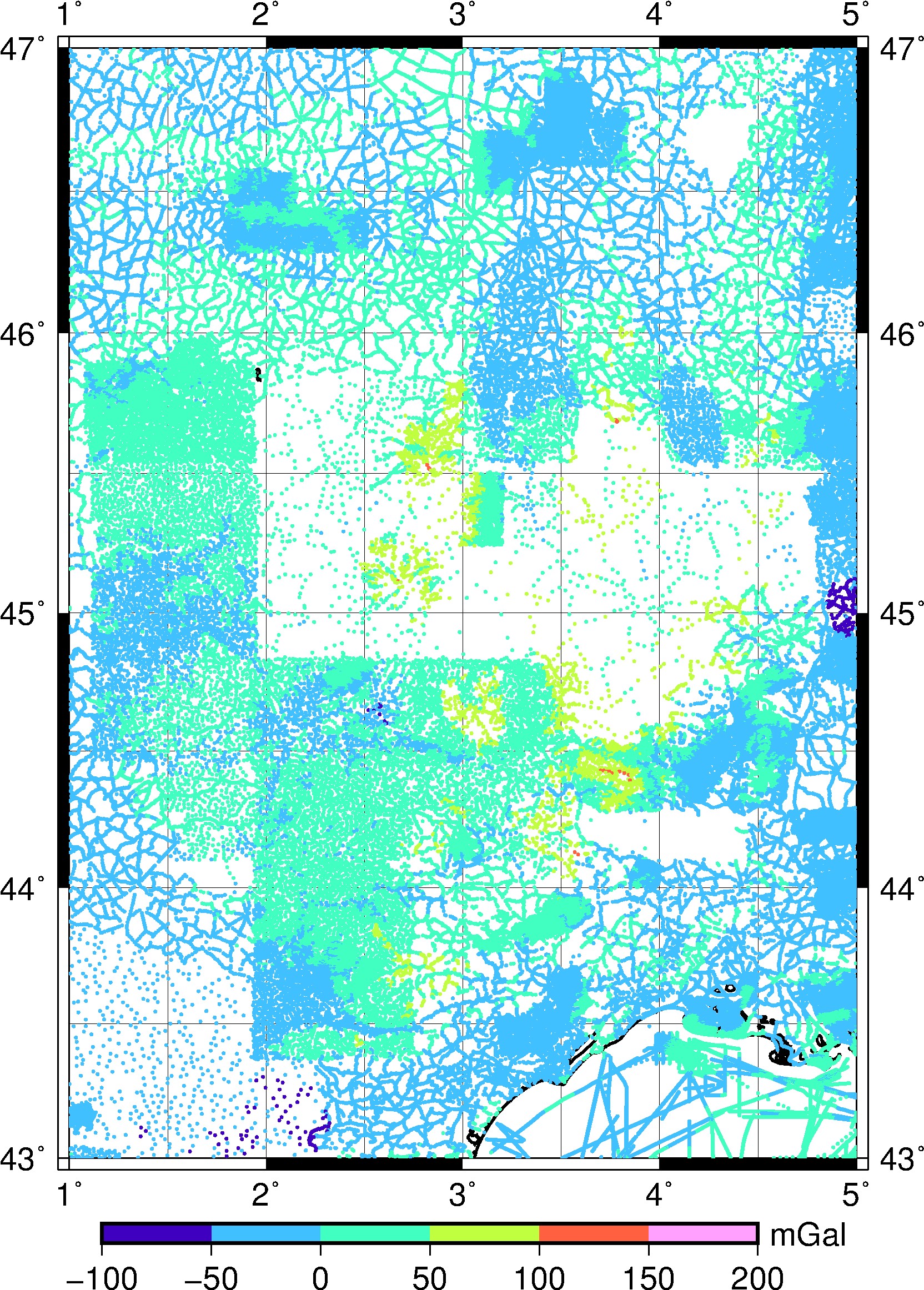}}
		\label{fig:Auv_BGI_dg_val}}
	\hfill\null
	\captionsetup{width=\linewidth}  
	\caption{Topography and gravity data distribution in the Massif Central area.}
	\label{fig:Auv}
\end{figure}

\begin{figure}
	\captionsetup{width=0.4\linewidth}  
	\centering
	\null\hfill
	\subfloat[Topography.]{%
		\includegraphics[clip=true, trim = 0 0 0 0,width=0.33\linewidth]{%
			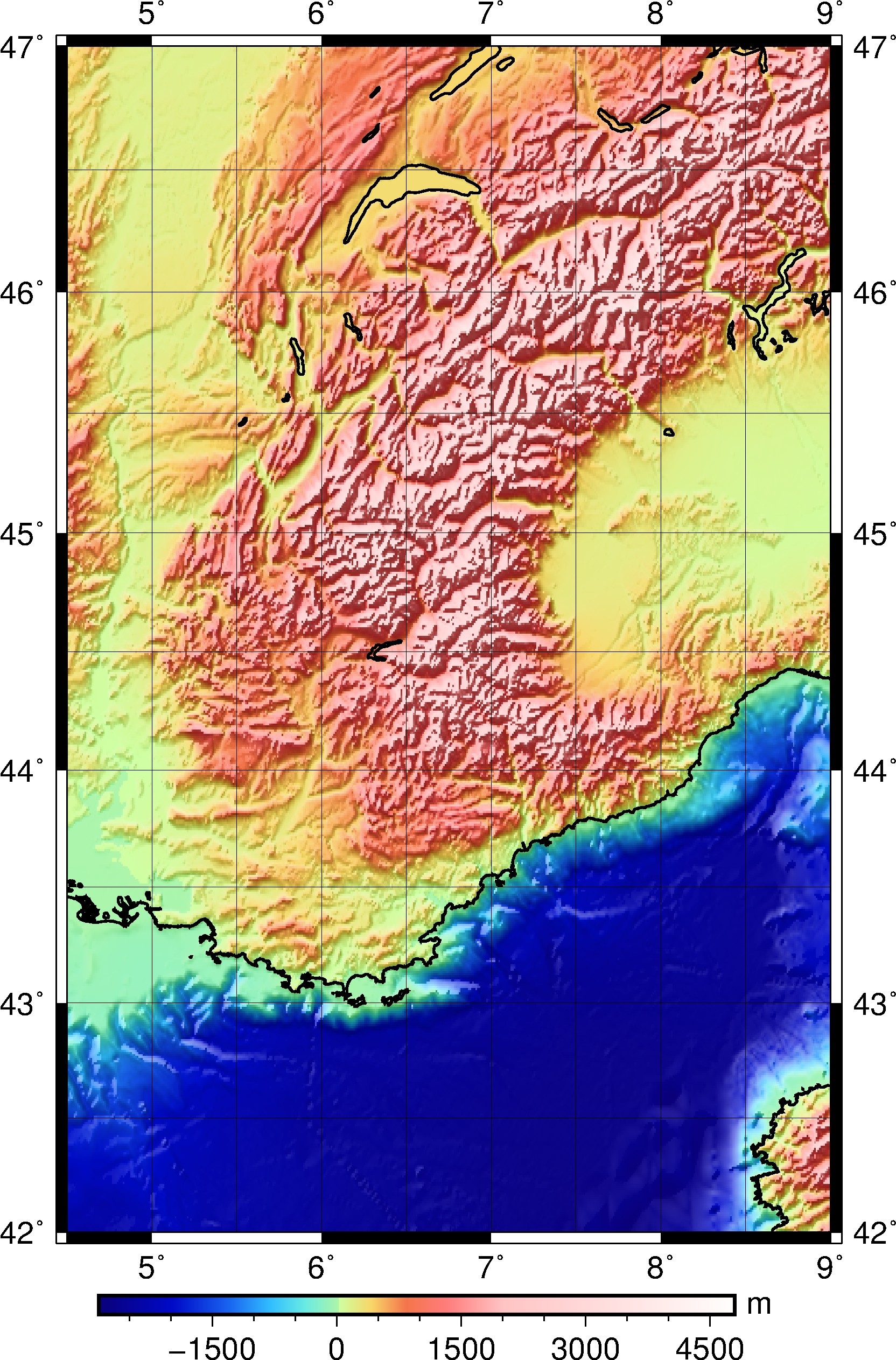}
		\label{fig:Alpes_topo}}
	\hfill
	\subfloat[Terrestrial and marine free-air gravity anomalies.]{%
		\raisebox{0\height}{\includegraphics[clip=true, trim = 0 0 0 0, width=0.33\linewidth]{%
				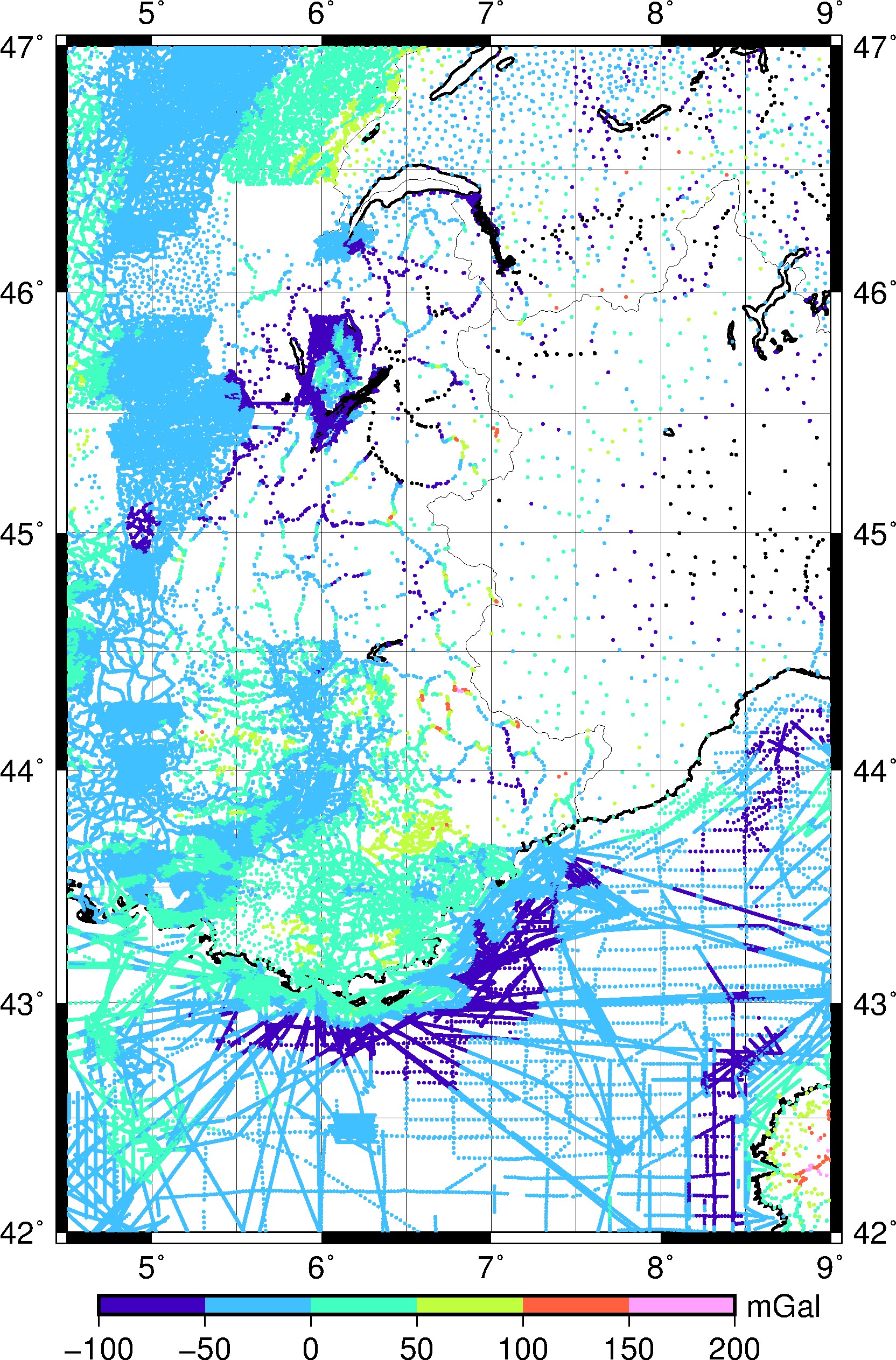}}
		\label{fig:Alpes_BGI_dg_val}}
	\hfill\null
	\captionsetup{width=\linewidth}  
	\caption{Topography and gravity data distribution in the Alps-Mediterranean area.}
	\label{fig:Alpes}
\end{figure}

Our study focuses on two different areas in France. 
The first region is the Massif Central located between \SIrange{43}{47}{\degree}N and \SIrange{1}{5}{\degree}E, and consists of plateaus and low mountain range, see Figure~\ref{fig:Auv}.
The second target area, much more hilly and mountainous, is the French Alps with a portion of the Mediterranean Sea located at the limit of different countries and bounded by \SIrange{42}{47}{\degree}N and \SIrange{4.5}{9}{\degree}E, see Figure~\ref{fig:Alpes}.
Topography is obtained from the 30~m digital elevation model over France by IGN, completed with \citet{Smith_1997aa} bathymetry and SRTM data.


Available surface gravity data in these areas, from the BGI (International Gravimetric Bureau), are shown in Figures~\ref{fig:Auv_BGI_dg_val}--\ref{fig:Alpes_BGI_dg_val}. Note that the BGI gravity data values are not used in this study, but only their spatial distribution in order to generate realistic distribution in the synthetic tests.
%
%
In these figures, it is shown that the gravity data are sparsely distributed: the plain is densely surveyed while the mountainous regions are poorly covered because they are mostly inaccessible by the conventional gravity survey. The range of free-air gravity anomalies~\citep[see e.g][]{Moritz_1980aa,Sanso_2013aa} which are quite large reflects the complex structure of the gravity field in these regions, which means that the gravitational field strength varies greatly from place to place at high-resolution. The scarcity of gravity data in the hilly regions is thus a major limitation in deriving accurate high-resolution geopotential model. 


%
%
%
%
%
%
%
%

\subsection{High resolution synthetic data}
\label{ssec:data_synt}

Here, we present the way to simulate our synthetic gravity disturbances~$\Xref{\grdist}$ and disturbing potentials~$\Xref{T}$ by subtracting the gravity field long and short wavelengths influence of a high-resolution global geopotential model.

The generation of the synthetic data~$\Xref{\grdist}$ and~$\Xref{T}$ at the Earth's topographic surface was carried out, in ellipsoidal approximation, with the Fortran program GEOPOT\footnote{http://www.ngs.noaa.gov/GEOID/RESEARCH\_SOFTWARE/research\_software.html} \citep{Smith_1998aa} of the National Geodetic Survey (NGS). \textcolor{black}{This program allows to compute gravity field related quantities at given locations using a geopotential model and additional informations such as parameters of the ellipsoidal normal field, tide system.} The ellipsoidal normal field is defined by the parameters of the geodetic reference system GRS80 \citep{Moritz_1980ab}. 
As input, we used the static global gravity field model \linebreak[2] EIGEN-6C4 \citep{Forste_2014aa}. It is a combined model up to degree and order (d/o) \num{2190} containing satellite, altimetry, terrestrial gravity and elevation data. By using the spherical harmonics (SH) coefficients up to d/o~\num{2000}, it allows us to map gravity variations down to 10~km resolution. \textcolor{black}{Thus, these synthetic data do not represent  the full geoid signal. The choice is motivated by the fact that at a centimetric level of accuracy, we expect large benefit from clocks at wavelengths $\ge 10$~km}.

\begin{figure}[!htbp]
	\captionsetup{width=0.6\linewidth}  
	\centering
	\includegraphics[clip=true, trim = 0 0 0 0, width=0.45\linewidth]{%
		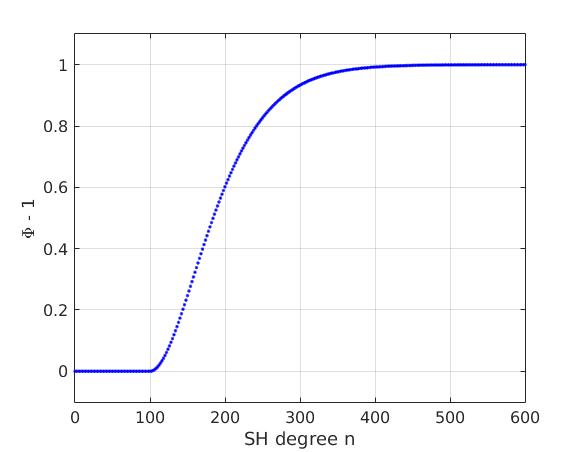}
	\caption{High pass filter based on a Poisson wavelet~$\Phi$ at order~$m=3$. The cutoff is~$n_{\textrm{cut}}=100$ and the wavelet scale is 0.03.}	
	\label{fig:Filtren100}
\end{figure}

Our objective is to study how clocks can advance knowledge of the geoid beyond the resolution of the satellites. 
In a first step, as illustrated in Figure~\ref{fig:Filtren100}, the long wavelengths of the gravity field covered by the satellites \textcolor{black}{and longer than the extent of the local area} are completely removed up to the degree~$n_{\textrm{cut}} = 100$ (200 km resolution). \textcolor{black}{This data reduction is necessary for the determination of \textcolor{black}{the} local covariance function in order to have centered data, or close to zero, as detailed in \cite{Knudsen_1987aa,Knudsen_1988aa}.} Between degree~101 and 583, the gravity field is progressively filtered using 3~Poisson wavelets spectra \citep[][]{Holschneider_2003aa}, while its full content is preserved above degree~583. In this way we realize a smooth transition between the wavelengths covered by the satellites and those constrained from the surface data.

%

%
%
%

\begin{figure}[!htbp]
	\captionsetup{width=0.35\linewidth}  
	\centering
	\null\hfill
	\subfloat[$\Xref{\grdist}$]{%
		\includegraphics[clip=true, trim = 0 0 0 0, width=0.33\linewidth]{%
			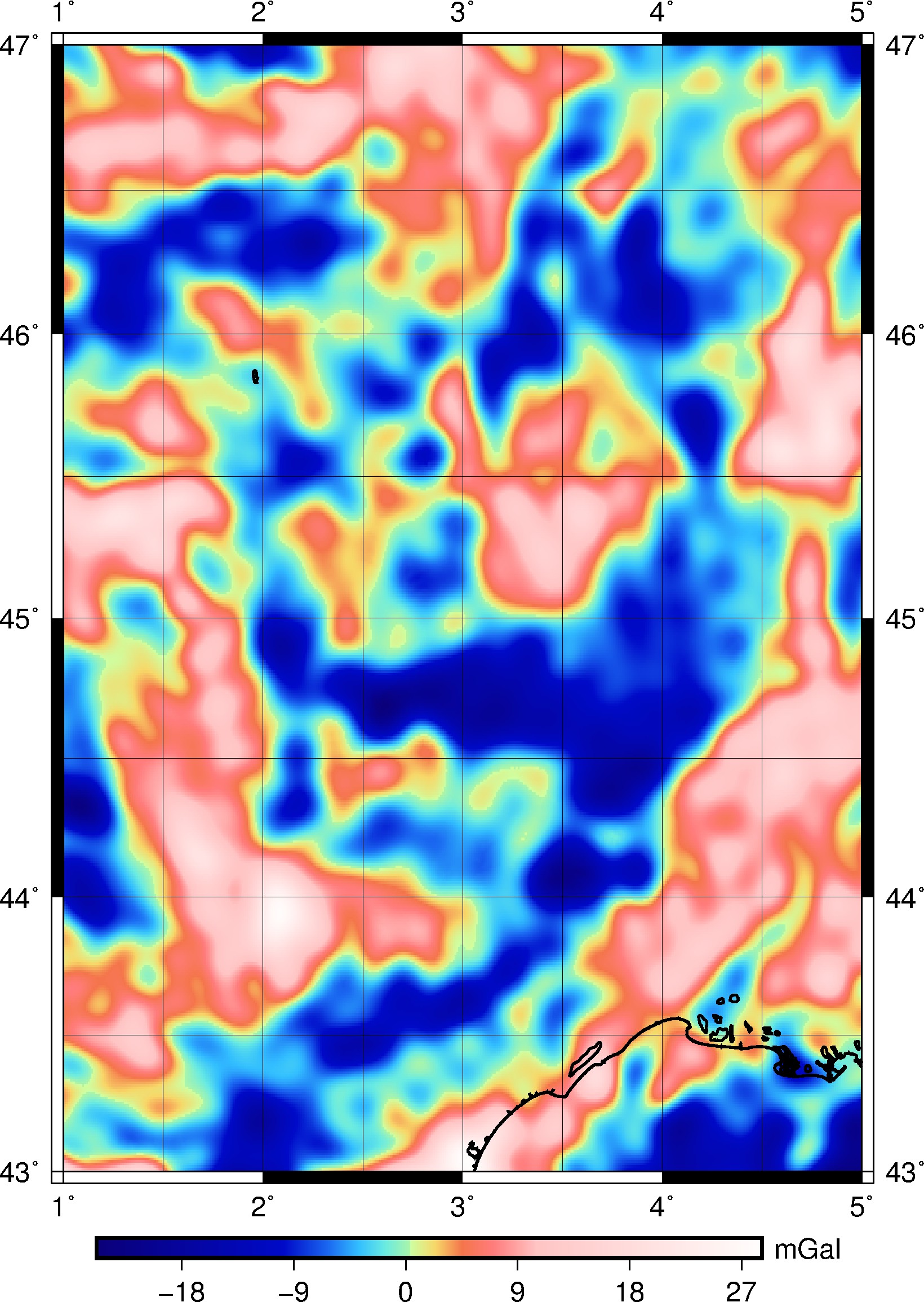}
		\label{fig:Auv_dg_geopot}}
	\hfill
	\subfloat[$\Xref{T}$]{%
		\raisebox{0\height}{\includegraphics[clip=true, trim = 0 0 0 0, width=0.33\linewidth]{%
				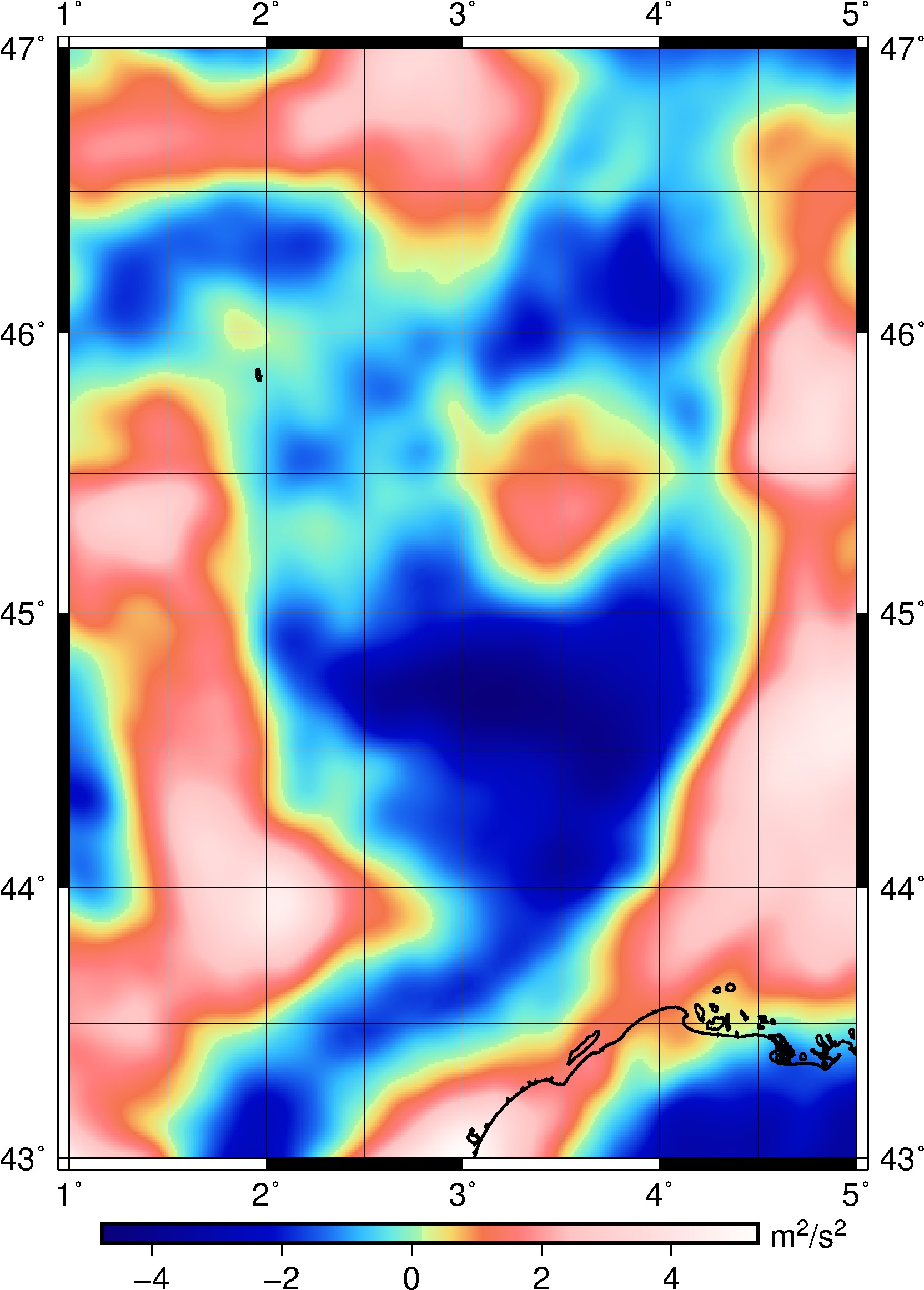}}
		\label{fig:Auv_T_geopot}}
	\hfill\null
	\captionsetup{width=\linewidth}  
	\caption{Synthetic reference fields of gravity disturbances~$\Xref{\grdist}$ and disturbing potential~$\Xref{T}$ in the Massif Central area. Anomalies are computed at the Earth's topographic surface from the EIGEN-6C4 model up to d/o~\num{2000} after removal of the low and high frequencies of the gravity field.}
	\label{fig:Auv_Tdg_geopot}
\end{figure}

\begin{figure}[!htbp]
	\captionsetup{width=0.35\linewidth}  
	\centering
	\null\hfill
	\subfloat[$\Xref{\grdist}$]{%
		\includegraphics[clip=true, trim = 0 0 0 0, width=0.33\linewidth]{%
			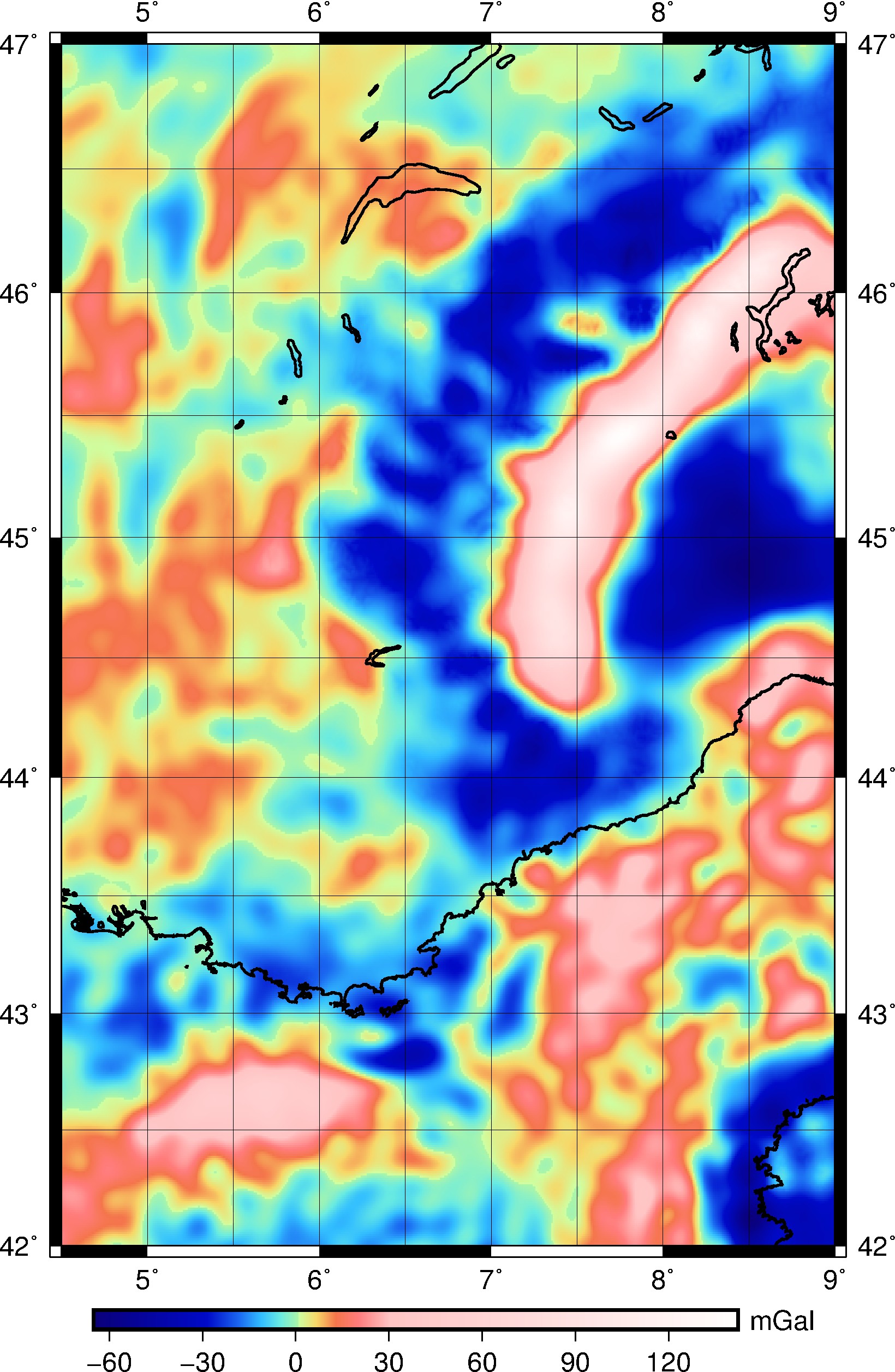}
		\label{fig:Alpes_dg_geopot}}
	\hfill
	\subfloat[$\Xref{T}$]{%
		\raisebox{0\height}{\includegraphics[clip=true, trim = 0 0 0 0, width=0.33\linewidth]{%
				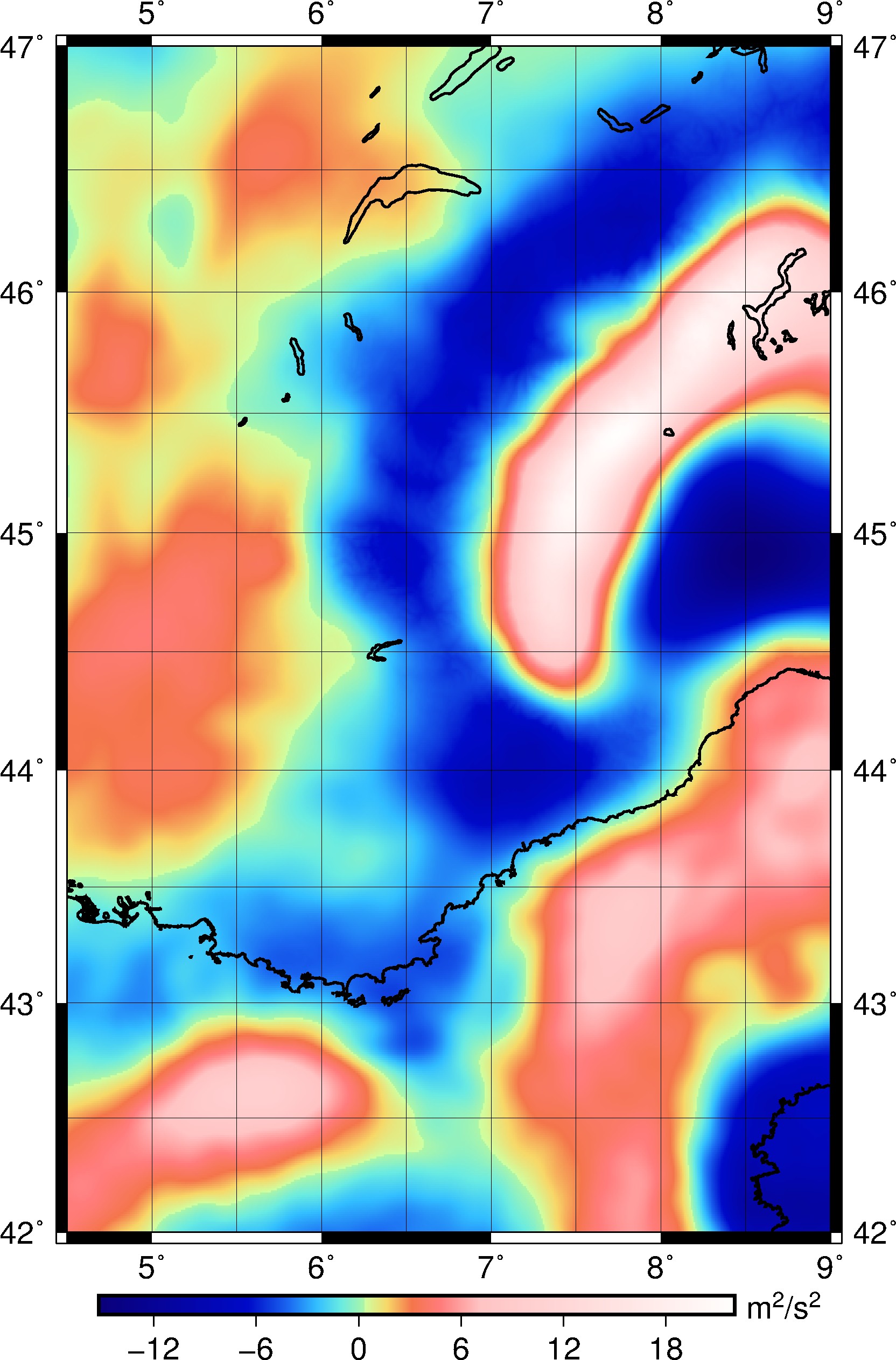}}
		\label{fig:Alpes_T_geopot}}
	\hfill\null
	\captionsetup{width=\linewidth}  
	\caption{Synthetic reference fields of gravity disturbances~$\Xref{\grdist}$ and disturbing potential~$\Xref{T}$ in the Alps-Mediterranean area. Anomalies are computed at the Earth's topographic surface from the EIGEN-6C4 model up to d/o~\num{2000} after removal of the low and high frequencies of the gravity field.}
	\label{fig:Alpes_Tdg_geopot}
\end{figure}

To subtract the terrain effects included in EIGEN-6C4, we used the topographic potential model dV\_ELL\_RET2012 \citep{Claessens_2013aa} truncated at d/o~\num{2000}. Complete up to d/o~\num{2160}, this model provides in ellipsoidal approximation the gravitational attraction due to the topographic masses anywhere on the Earth’s surface. The results of this data reduction yields to the reference fields~$\Xref{\grdist}$ and~$\Xref{T}$ for both regions, shown in Figures~\ref{fig:Auv_Tdg_geopot}--\ref{fig:Alpes_Tdg_geopot}. 

\textcolor{black}{The Figures \ref{fig:Auv_Tdg_geopot}--\ref{fig:Alpes_Tdg_geopot} show the different characteristics of the residual field in these two regions. The residual anomalies have smaller amplitudes in the Massif Central area when compared to the Alps. In addition, the presence of high mountains on part of the latter zone results in an important spatial heterogeneity of the residual gravity anomalies, with large signals also at intermediate resolutions.}

\section{Dataset selection and synthetic noise}
\label{sec:4}

\paragraph{\textbf{Gravimetric location points selection}.}

Our goal is to reproduce a realistic spatial distribution of the gravity points. The BGI gravity data sets contain hundreds of thousands points for the target regions (see Figure~\ref{fig:Auv_BGI_dg_val}--\ref{fig:Alpes_BGI_dg_val}). In order to reduce the size of the problem and make it numerically more tractable, we build a distribution with no more than several thousand points from the original one.

\begin{figure}[!htbp]
	\captionsetup{width=0.35\linewidth}  
	\centering
	\null\hfill
	\subfloat[Massif Central: \num{4374} gravity data and \num{33}~potential data.]{%
		\raisebox{0\height}{\includegraphics[clip=true, trim = 0 0 0 0, height=0.30\paperheight]{%
				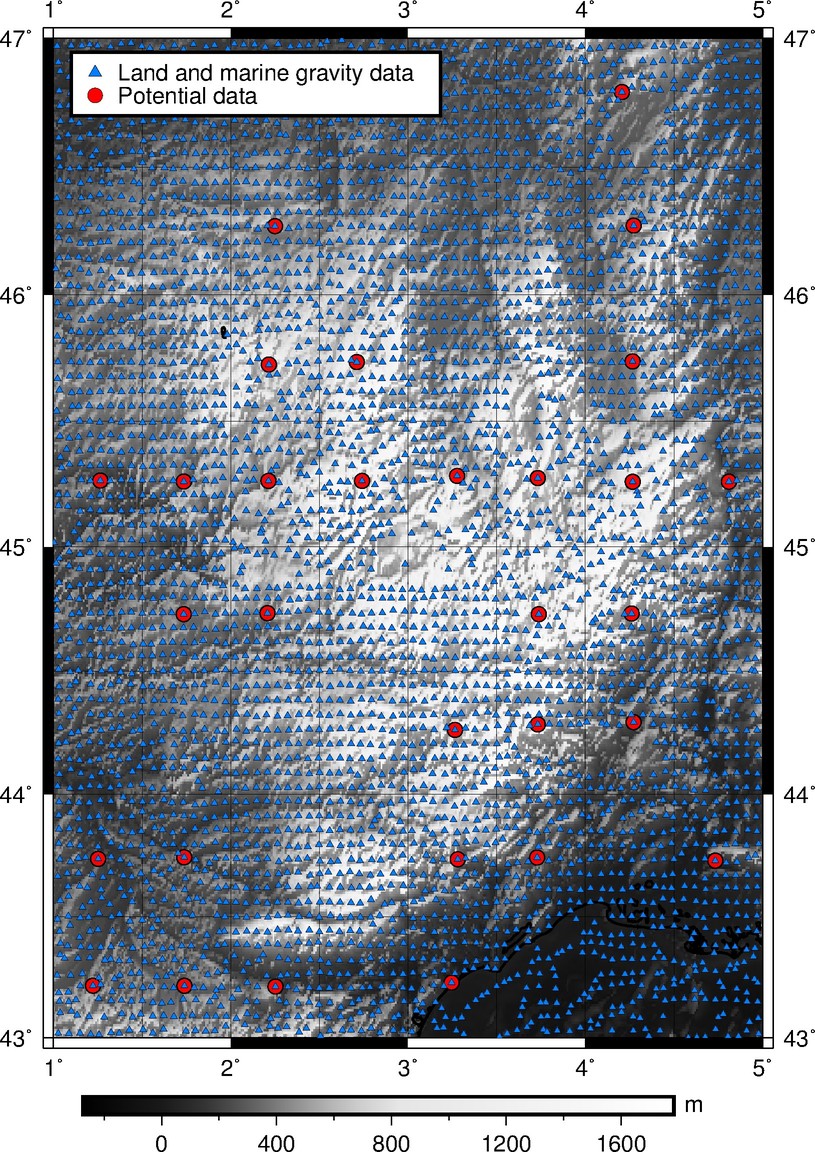}}
		\label{fig:Auv_Coverage_Tdg_33}}
	\hfill
	\subfloat[Alps: \num{4959} gravity data and \num{32}~potential data.]{%
		\includegraphics[clip=true, trim = 0 0 0 0, height=0.30\paperheight]{%
			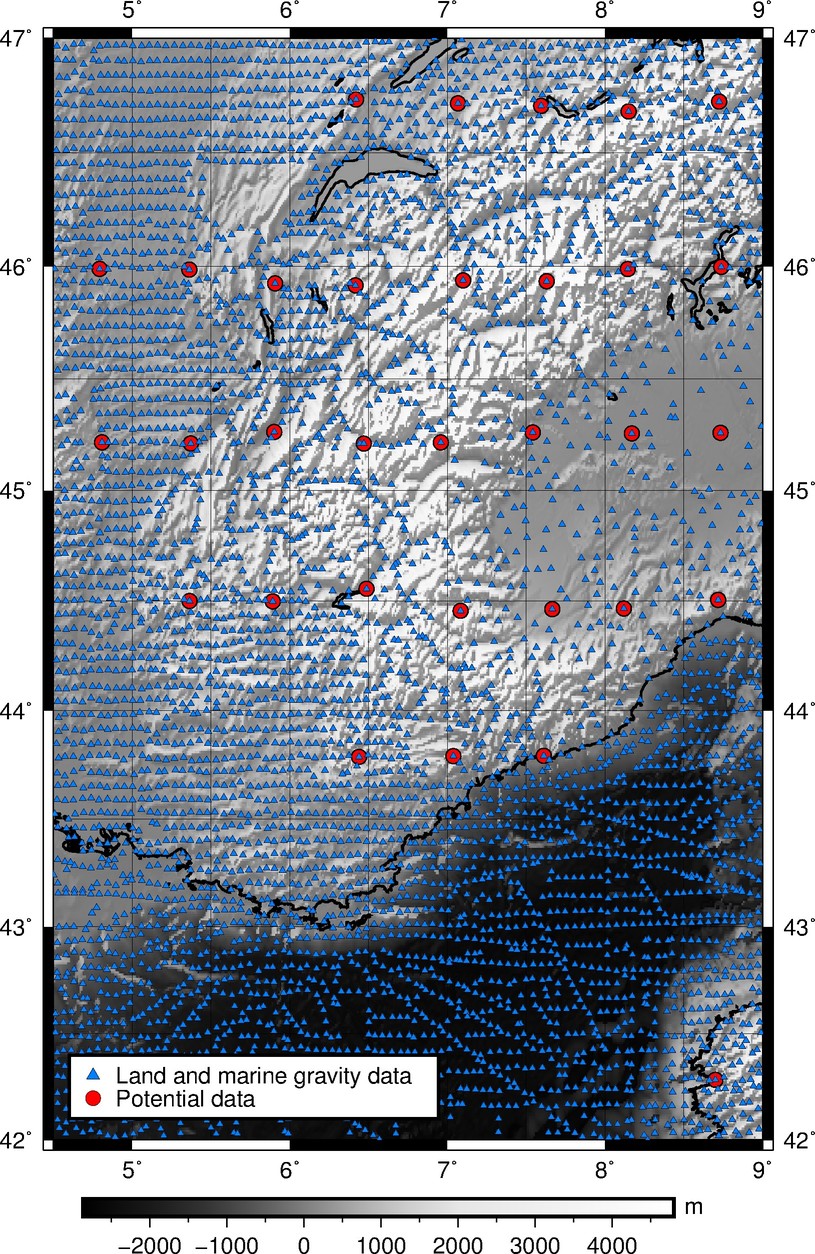}
		\label{fig:Alpes_Coverage_Tdg_32}}
	\hfill\null
	
	\captionsetup{width=\linewidth}  
	\caption{Distribution of the gravity and clock data used in the synthetic tests.}
	\label{fig:Coverage_Tdg}
\end{figure}

Starting from the spatial distribution of the BGI gravity data sets, a grid~$\Xref{\grdist}$ of~$N$~cells is built with a regular step of about 6.5~km. Each cell contains~$n_i$ points with~$i=\{1,2,\dots,N\}$. These~$n_i$ points are replaced by one point which location is given by the geometric barycenter of the~$n_i$ points, in the case that~$n_i>0$. If~$n_i=0$ then there is no point in the cell~$i$. Figures~\ref{fig:Coverage_Tdg} show the new distributions of gravimetric data for the Massif Central and the Alps regions; they have, respectively, \num{4374} and \num{4959} location points. These new spatial distributions reflect the initial BGI gravity data distribution but are be more homogeneous. They will be used in what follows.

%



\paragraph{\textbf{Chronometric location points selection}.}

We choose to put clock measurements only where existing land gravity data are located. Indeed, these data mainly follow the roads and valleys which could be accessible for a clock comparison. Then, we use a simple geometric approach in order to put clock measurements in regions where the gravity data coverage is poor. Since the potential varies smoothly compared to the gravity field, a clock measurement is affected by masses at a larger distance than in the case of a gravimetric measurement. For that reason, a clock point will be able to constrain longer wavelengths of the geopotential than a gravimetric point.
%
%
\textcolor{black}{This is particularly interesting in areas poorly surveyed by gravity \textcolor{black}{measurement} networks.} Finally, in order to avoid having clocks too close to each other, we define a minimal distance~$d$ between them. We chose~$d$ greater than the correlation length of the gravity covariance function (in this work~$\lambda \sim 20$~km, see Table~\ref{tab:cov_stats}).

Here we give more details about our algorithm to select the clock locations:
\begin{enumerate} 
	\setlength{\itemsep}{0pt}
	\item First, we initialize the clock locations on the nodes of a regular grid~$\Xref{T}$ with a fixed interval~$d$. This grid is included in the target region at a setback distance of about 30~km from each edge (outside possible boundary effects);
	\item Secondly, we change the positions of each clock point to the position of the nearest gravity point from the grid~$\Xref{\grdist}$, located in cell~$i$ (see the previous paragraph); in cell~$i$ are located~$n_i$ points of the initial BGI gravity data distribution;
	\item Finally, we remove all the clock points located in cells where~$n_i > n_{max}$. This is a simple way to keep only the clock points located in areas with few gravimetric measurements.
\end{enumerate}
This method allows to simulate different realistic clock measurement coverages by changing the values of~$d$ and~$n_{max}$. The number of clock measurements increases when the distance~$d$ decreases or when the threshold~$n_{max}$ increases, and vice versa. It is also possible to obtain different spatial distributions but the same number of clock measurements for different sets of~$d$ and~$n_{max}$.

In Figure~\ref{fig:Coverage_Tdg}, we propose an example of clock coverage used hereafter for both target regions with 32 and 33 clock locations, respectively, in the Massif Central and the Alps, corresponding to~$\sim 0.7$ percent of the gravity data coverage. For the chosen distributions, the value of~$d$ is about 60~km and~$n_{max} = 15$.



\paragraph{\textbf{Synthetic measurements simulation}.} For each data point, the synthetic values of~$\grdist$ and~$T$ are computed by applying the data reduction presented in Section~\ref{ssec:data_synt}. It is important to note that the location points of the simulated data~$T$ are not necessarily at the same place than the estimated data~$\Xref{T}$.

A Gaussian white noise model is used to simulate the instrumental noise of the measurements. We chose, for the main tests in the next section, a standard deviation~$\sigma_{\grdist}=1$~mGal for the gravity data and~$\sigma_{T}=0.1$~m$^2$/s$^2$ for the potential data. In terms of geoid height, the latter noise level is equivalent to 1~cm. Other tests with different noise levels are discussed in Section~\ref{sec:6}.

\section{Numerical results}
\label{sec:5}

In this section, we present our numerical results showing the contribution of clock data in regional recovery of the geopotential from realistic data points distribution in the Massif Central and the Alps. The reconstruction of the disturbing potential is realized from the synthetic measurements~$\grdist$ and~$T$, and by applying the Least-Squares Collocation (LSC) method.

\paragraph{\textbf{Planar Least-Squares Collocation}.}
\label{ssec:LSC}

The LSC method, described in \cite{Moritz_1972aa, Moritz_1980aa}, is a suitable tool in geodesy to combine heterogeneous data sets in gravity field modelling. Assuming that the measured values are linear functionals of the disturbing potential~$T$, this approach allows us to estimate any gravity field parameter based on~$T$ from many types of observables. 

Consider~$\vec{l} = [ \vec{l}_T, \vec{l}_{\grdist} ] = l_k$ a data vector composed by~$p$ data~$T$ and~$q$ data~$\grdist$, affected by measurement errors~$\epsilon_k$, with~$k=\{1,2,\ldots,p+q\}$. The estimation of the disturbing potential~$\widetilde{T}_P$ at point~$P$ from the data~$\vec{l}$ can be performed with the relation
\begin{align}
\label{eq:def_LSC}
\widetilde{T_P} & = {}
\matr{C}_{T_P,l}^{\intercal} \cdot \matr{C}_{n,n}^{-1} \cdot \vec{l} 
\\
\matr{C}_{n,n} & = {}
\matr{C}_{l,l} + \omega\,\matr{C}_{\epsilon,\epsilon}
\end{align}
with~$\matr{C}_{l,l}$ the covariance matrix of the measurement vector~$\vec{l}$,~$\matr{C}_{\epsilon,\epsilon}$ the covariance matrix of the noise,~$\matr{C}_{T_P,l}$ the cross covariance matrix between the estimated signal~$T_P$ and the data~$\vec{l}$, and~$\omega$ the Tikhonov regularization factor \citep{Neyman_1979aa}, also called weight factor.

In practice, the data~$\vec{l}$ are synthesized as described in Sections~\ref{sec:3} and~\ref{sec:4}. Therefore, the measurement noise is known to be a Gaussian white noise. Noise and signal (errorless part of~$l_k$) are assumed to be uncorrelated, and the covariance matrix of the noise can be written as
\begin{equation}
\label{eq:def_Cee}
\matr{C}_{\epsilon,\epsilon} = {}
\left[
\begin{array}{cc}
\matr{I}_p \cdot\sigma_T^2 & 0 \\ 
0 & \matr{I}_q \cdot\sigma_{\grdist}^2
\end{array}
\right]
\end{equation}
with~$\matr{I}_n$ the identity matrix of size~$n$. 

Because~$\matr{C}_{l,l}$ can be very ill-conditioned, the matrix \eqref{eq:def_Cee} plays an important role in its regularization before inversion, since positive constant values are added to the elements of its main diagonal. To avoid any iterative process to find an optimum value of~$\omega$ in case where this matrix~$\matr{C}_{l,l}$ is not definite positive, we chose to fix the weight factor~$\omega = 1$ and to apply a singular value decomposition (SVD) to pseudo-inverse the matrix. As shown in \citep{Rummel_1979aa}, these two approaches are similar.

\paragraph{\textbf{Estimation of the covariance function}.}
\label{ssec:EFC}

Implementation of the collocation method requires to compute the covariance matrices~$\matr{C}_{T_{P},l}$ and~$\matr{C}_{l,l}$. This step has been carried out using a logarithmic spatial covariance function from \citep{Forsberg_1987aa}, see Appendix~\ref{anx:defFCF}. This stationary and isotropic model is well-adapted to our analysis. Indeed, it provides the auto-covariances (ACF) and cross-covariances (CCF) of the disturbing potential~$T$ and its derivatives in 3~dimensions with simple closed-form expressions.


The spatial correlations of the gravity field are analyzed with the program GPFIT \citep{Forsberg_2008aa}. The variance~$C_0$ is directly computed from the gravity data on the target area, and the parameters~$\alpha$ and~$\beta$ (see Appendix~\ref{anx:defFCF}) are estimated by fitting the a priori covariance function to the empirical ACF of the gravity disturbances~$\grdist$.

\begin{figure}[!htbp]
	\captionsetup{width=\linewidth}  
	\centering
	\null\hfill
	\subfloat[Massif Central.]{%
		\raisebox{0\height}{\includegraphics[clip=true, trim = 0 0 0 0, width=0.35\linewidth]{%
				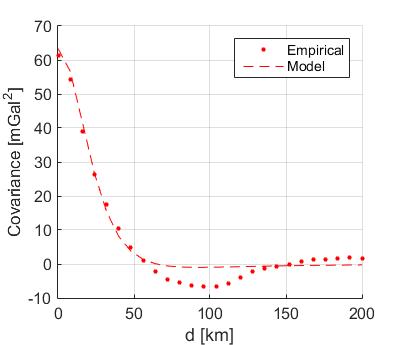}}
		\label{fig:Auv_cov_dg}}
	\hfill
	\subfloat[Alps-Mediterranean.]{%
		\includegraphics[clip=true, trim = 0 0 0 0, width=0.35\linewidth]{%
			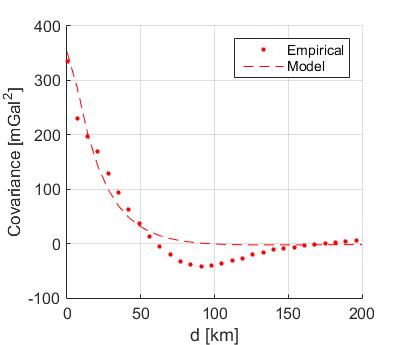}
		\label{fig:Alpes_cov_dg}}
	\hfill\null
	\captionsetup{width=\linewidth}  
	\caption{Empirical and best fitting covariance function of the ACF of~$\grdist$. Values of the parameters are given in Table~\ref{tab:cov_stats}.}
	\label{fig:Covariogram}
\end{figure}

Results of the optimal regression analysis for both regions are given in Figure~\ref{fig:Covariogram} and Table~\ref{tab:cov_stats}. \textcolor{black}{The estimated covariance \textcolor{black}{models} reflect the different characteristics of the gravity signals in the two areas and the data sampling, which is less dense in high relief areas. Finally, the gravity \textcolor{black}{anomaly} covariances show similar correlation lengths, with a larger variance for the case of the Alps; their shapes, however, slightly differ, with a broader spectral coverage for the Alps.}

\begin{table}[!htbp] 
	
	\sisetup{
		table-number-alignment = left,
		table-figures-integer = 1, 
		table-figures-decimal = 0,
		table-auto-round,
		table-sign-mantissa,
		table-sign-exponent,
		exponent-product =  \times, 
		output-exponent-marker 
	} 
	\captionsetup{width=0.9\linewidth}  
	\centering
	\resizebox{0.8\linewidth}{!}{%

	\begin{footnotesize}
		\begin{tabular}{
				*{1}{K{6em}}
				*{5}{S[table-format = -1.1e0]} |
				*{1}{S[table-format = -1.1e0]}
			} 

			\toprule 
			
			\multicolumn{1}{l}{\textrm{Area}} 
			& \multicolumn{1}{c}{\textrm{Nb data}} 
			& \multicolumn{1}{c}{$\mu$ [mGal]}
			& \multicolumn{1}{c}{$C_0$ [mGal$^2$]}
			& \multicolumn{1}{c}{$\alpha$ [km]}
			& \multicolumn{1}{c}{$\beta$ [km]}
			& \multicolumn{1}{c}{$\lambda$ [km]} \\ 
			
			\cmidrule(rl){1-1}
			\cmidrule(rl){2-6}
			\cmidrule(rl){7-7}
			
			\textrm{Massif Central}
			& 4374
			& 0.41
			& 63.4
			& 24
			& 15
			& 21
			\\
			\textrm{Alps--Med.}
			& 4959
			& 1.15
			& 352.5
			& 6
			& 47
			& 18
			\\

			\bottomrule 
		\end{tabular}
	\end{footnotesize}}
	\caption{Estimation of the auto covariance function parameters on the gravity data~$\grdist$ using the logarithmic model from \cite{Forsberg_1987aa} with,~$\mu$ the mean,~$C_0$ the variance,~$\alpha$ and~$\beta$ respectively a shallow and a compensating depth parameter. Here,~$\lambda$ is the correlation length defined as the distance at which the covariance is half of the variance.}
	\label{tab:cov_stats}
\end{table}

Knowing the parameter values of the covariance model, we can now estimate the potential anywhere on the Earth's surface. 
%

\paragraph{\textbf{Contribution of clocks.}}

The contribution of clock data in the potential recovery is evaluated by comparing the residuals of two solutions to the reference potential on a regular grid interval of~10~km. The first solution corresponds to the errors between the estimated potential model computed solely from gravity data and the potential reference model, while the second solution uses combined gravimetric and clock data. To avoid boundary effects in the estimated potential recovery, a grid edge cutoff of 30~km has been removed in the solutions.

\begin{figure}[!htbp]
	\captionsetup{width=0.8\linewidth}  
	\centering
	\null\vfill
	\subfloat[Without clock data.]{%
		\raisebox{0\height}{\includegraphics[clip=true, trim = 30 0 30 0, width=0.65\linewidth]{%
				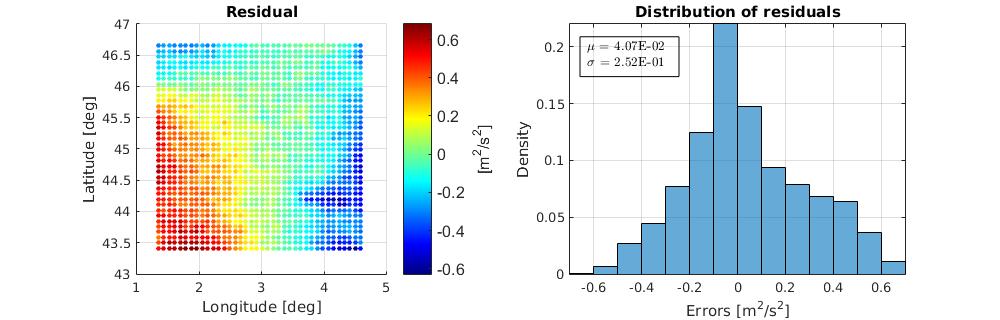}}
		\label{fig:Auv_EMC_Tdg_from_T0dg_4art}}
	\vfill
	\subfloat[With clock data.]{%
		\includegraphics[clip=true, trim = 30 0 30 0, width=0.7\linewidth]{%
			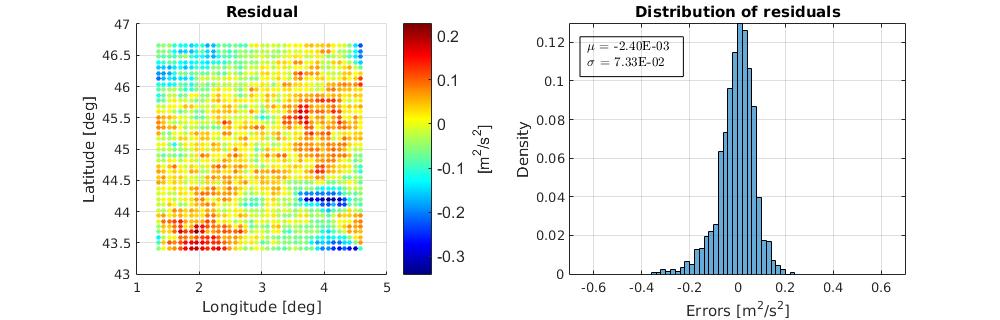}
		\label{fig:Auv_EMC_Tdg_from_T33dg_56_15_4art}}
	\vfill\null
	
	\caption{Accuracy of the disturbing potential~$T$ reconstruction on a regular 10-km step grid in Massif Central, obtained by comparing the reference model and the reconstructed one. In Figure (a), the estimation is realized from the \num{4374} gravimetric data~$\delta g$ only, and in Figure (b) by adding 33 potential data~$T$ to the gravity data.}
	\label{fig:Auv_T_recovery}
\end{figure} 
\begin{figure}[!htbp]
	\captionsetup{width=0.8\linewidth}  
	\centering
	\null\vfill
	\subfloat[Without clock data.]{%
		\raisebox{0\height}{\includegraphics[clip=true, trim = 30 0 30 0, width=0.65\linewidth]{%
				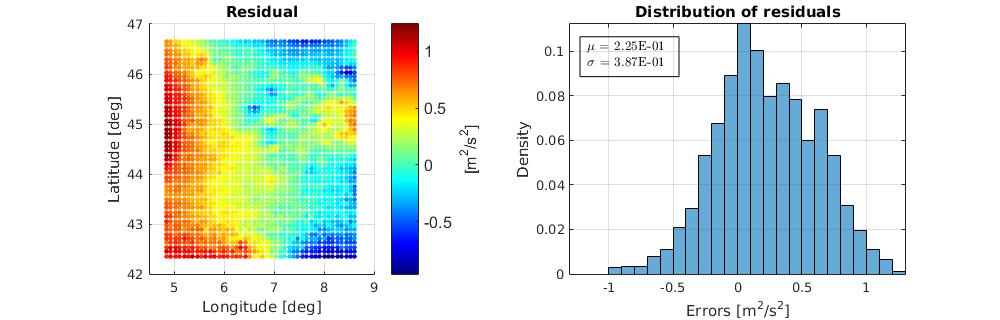}}
		\label{fig:Alpes_EMC_Tdg_from_T0dg_4art}}
	\vfill
	\subfloat[With clock data.]{%
		\includegraphics[clip=true, trim = 30 0 30 0, width=0.7\linewidth]{%
			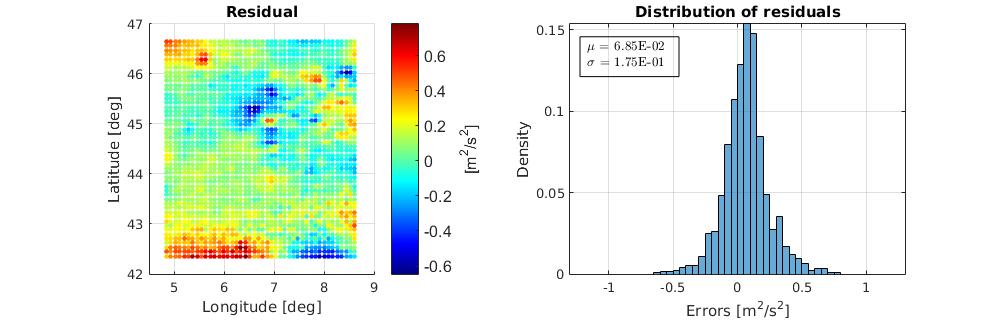}
		\label{fig:Alpes_EMC_Tdg_from_T32dg_50_3_4art}}
	\vfill\null
	
	\caption{Accuracy of the disturbing potential~$T$ reconstruction on a regular 10-km step grid in Massif Central, obtained by comparing the reference model and the reconstructed one. In Figure (a), the estimation is realized from the \num{4959} gravimetric data~$\delta g$ only, and in Figure (b) by adding 32 potential data~$T$ to the gravity data.}
	\label{fig:Alpes_T_recovery}
\end{figure} 

For the Massif Central region, the disturbing potential is estimated with a bias~$\mu_T \approx 0.041$~\mcpsc (4.1~mm) and a rms~$\sigma_T \approx 0.25$~\mcpsc (2.5~cm) using only the \num{4374}~gravimetric data, see Figure~\ref{fig:Auv_EMC_Tdg_from_T0dg_4art}. When we now reconstruct~$T$ by adding the 33~potential measurements to the gravimetric measurements, the bias is improved by one order of magnitude ($\mu_T \approx -0.002$~\mcpsc~or~$-0.2$~mm) and the standard deviation by a factor~3 ($\sigma_T \approx 0.07$~\mcpsc~or 7~mm), see Figure~\ref{fig:Auv_EMC_Tdg_from_T33dg_56_15_4art}. 


For the Alps, Figure~\ref{fig:Alpes_T_recovery}, the potential is estimated with a bias~$\mu_T \approx 0.23$~\mcpsc (2.3~cm) and a standard deviation~$\sigma_T \approx 0.39$~\mcpsc (3.9~cm) using only the \num{4959}~gravimetric data. When adding the 32~potential measurements, we note that the bias is improved by a factor 4 ($\mu_T \approx -0.069$~\mcpsc or $-6.9$~mm) and the standard deviation by a factor~2 ($\sigma_T \approx 0.18$~\mcpsc~or 1.8~cm). 

\textcolor{black}{It can be noticed that the residuals in both areas differ. This results from the covariance function that is less well modeled when the data survey has large spatial gaps.} It should also be stressed that a trend appears in the reconstructed potential with respect to the original one when no clock data are added in both regions. This effect is discussed in Section~\ref{sec:6}.


\section{Discussion}
\label{sec:6}

\paragraph{\textbf{Effect of the number of clock measurements.}}
\label{ssec:nbclockeffect}

\begin{figure}[!htbp]
	\captionsetup{width=\linewidth}  
	\centering
	\null\hfill
	\subfloat[Massif Central area.]{%
		\raisebox{0\height}{\includegraphics[clip=true, trim = 0 0 0 0, width=0.45\linewidth]{%
				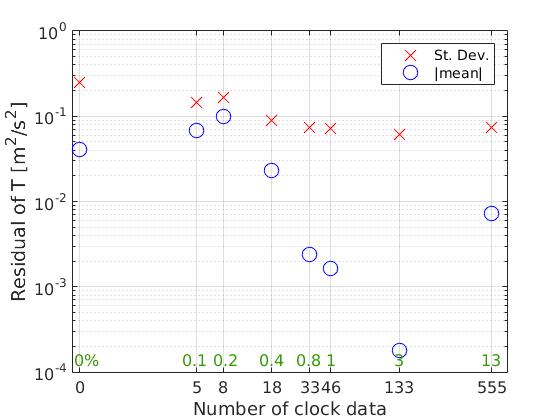}}
		\label{fig:Auv_Coverage_Effect}}
	\hfill
	\subfloat[Alps area.]{%
		\includegraphics[clip=true, trim = 0 0 0 0, width=0.45\linewidth]{%
			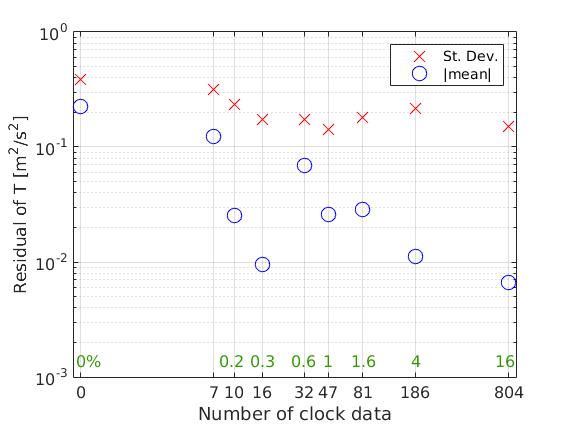}
		\label{fig:Alpes_Coverage_Effect}}
	\hfill\null
	\caption{Performance of the potential reconstruction (expressed by the standard deviations and mean of differences between the original potential on the regular grid and the reconstructed one) wrt the number of clocks. In green: number of clock data in terms of percentage of~$\grdist$ data.}
	\label{fig:Effect_nb_clock}
\end{figure}

Figure~\ref{fig:Effect_nb_clock} shows the influence of the number of clock data in the potential recovery, and thereby of their spatial distribution density. We vary the number and distribution of clock data by changing the mesh grid size~$d$, which represents the minimum distance between clock data points (see Section~\ref{sec:4}). The particular cases shown in detail in Section~\ref{sec:5} are included. We characterize the performance of the potential reconstruction by the standard deviation and mean of the differences between the original potential on the regular grid and the reconstructed one. When increasing the density of the clock network, the standard deviation of the differences tends toward the centimeter level, for the Massif Central case, and the bias can be reduced by up to 2 orders of magnitude. Note that we have not optimized the clock locations such as to maximize the improvement in potential recovery. The chosen locations are simply based on a minimum distance and a maximum coverage of gravity data (c.f. Section~\ref{sec:4}). An optimization of clock locations would likely lead to further improvement, but is beyond the scope of this work and will be the subject of future studies.  

Moreover, the results indicate that it is not necessary to have a large number of clock data to improve the reconstruction of the potential. We can see that only a few tens of clock data, i.e. less than 1~percent of the gravity data coverage, are sufficient to obtain centimeter level standard deviations and large improvements in the bias. When continuing to increase the number of clock data the standard deviation curve seems to flatten at the cm level.


\vspace{1cm}

\begin{figure}[!htbp]
	\captionsetup{width=\linewidth}  
	\centering
	\null\hfill
	\subfloat[Absolute value of the mean.]{%
		\raisebox{0\height}{\includegraphics[clip=true, trim = 0 0 0 0, width=0.45\linewidth]{%
				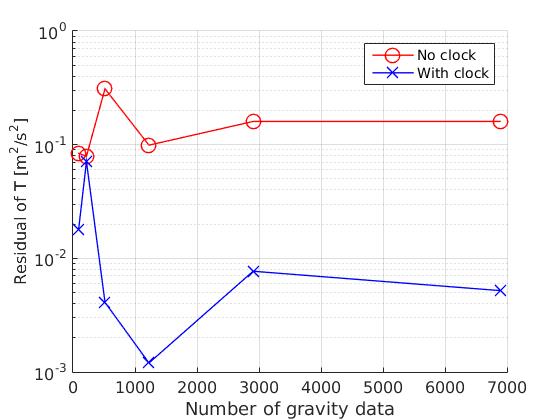}}
		\label{fig:Test_effet_densite_gravi_mean}}
	\hfill
	\subfloat[Rms.]{%
		\includegraphics[clip=true, trim = 0 0 0 0, width=0.45\linewidth]{%
			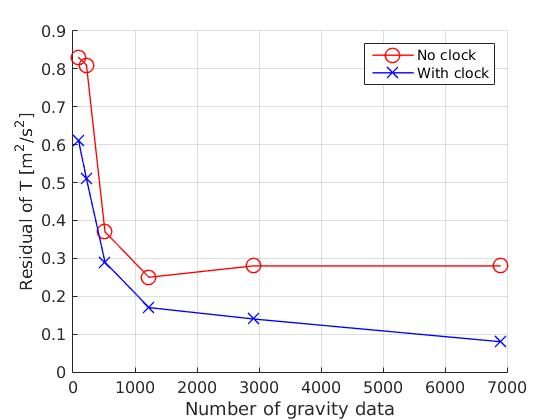}
		\label{fig:Test_effet_densite_gravi_sigma}}
	\hfill\null
	\caption{Effect of the number of gravity data \textcolor{black}{combined with 38 clock data} on the disturbing potential recovery in the Massif Central region. Panel (a): absolute value of the mean of the residuals of $T$; panel (b): the rms. The noise of the measurements is 1~mGal for $\grdist$ and $0.1~\mcpsc$ for $T$. Note that for each coverage of gravity data, a new covariance model is fitted on the empirical covariance model.}
	\label{fig:Test_effet_densite_gravi}
\end{figure}

\paragraph{\textbf{Effect of the number of gravity measurements.}}
\label{ssec:nbgraveffect}
\textcolor{black}{
We have performed numerical tests in order to study the influence of the density of gravity measurements on the reconstructed disturbing potential, with or without clocks. We take the case of the Massif Central region, and set-up simulations where the clock coverage is fixed (either no clocks, or 38 clocks at fixed locations where we also have gravity \textcolor{black}{data}). Then, we progressively increase the spatial resolution of the gravity data, from 91 to 6889 points, and evaluate as before the quality of the potential reconstruction with or without clocks. Here, in contrast with the tests presented in the previous section, the gravity points are randomly generated from a complete 5-km step grid.
Figure~\ref{fig:Test_effet_densite_gravi} shows the results of these tests. If we compare the rms values between configurations where we add clocks or not, we observe that the behavior of the results is globally similar, and improved with clocks. The interpolation error due to a too low resolution of the gravity data with respect to scales of the field variations predominates when we have less than $\sim$1500~gravity measurements, leading to large rms values even with clocks. Above this number, the large-scale reconstruction errors significantly contribute to the rms of residuals, explaining that the rms further decreases only when clocks are added. Looking at the bias between the reconstructed and original potential, we can see that it is poorly dependent on the number of gravity data in the tests without clocks. It probably reflects the fact that these data are more sensitive to the smaller scale components of the gravity potential. When we add clocks, the improvement on the bias is always important, which is consistent with the fact that the higher sensitivity of clocks to the longer wavelengths of the field reduces significantly the trend from the modelling error.}

\paragraph{\textbf{Covariance function consistency.}}
\label{ssec:syntdataeffect}

In Figures \ref{fig:Auv_EMC_Tdg_from_T0dg_4art} and \ref{fig:Alpes_EMC_Tdg_from_T0dg_4art}, a trend appears in the residuals, but disappears when gravimetric and clock data are combined. This is due to the fact that the covariance function does not have the same spectral coverage as the data generated from the gravity field model EIGEN-6C4. Indeed, the covariance function contains low frequencies while we have removed them for the synthetic data. Therefore, some low frequency content is present in the recovered potential. Whilst the issue could be avoided by using a covariance parametric model from which we can remove the low frequency content in a perfectly consistent way with the data generation (e.g. a closed-form Tscherning-Rapp model \citep{Tscherning_1974aa,Tscherning_1976aa}), it is not obvious that the corresponding results would allows realistic conclusions. Indeed, the spectral content of real surface observations, after removal of lower frequencies from a global spherical harmonics model, may still retain some {unknown} low frequencies. As consequence, it is not {obvious to match} to that of a single covariance function, while perfect consistency can only be achieved from synthetic data. We chose to keep this mismatch, thereby investigating the interest of clocks for high-resolution geopotential determination when our prior knowledge on the surface data signal and noise components is not perfect. More detailed studies on this issue are considered beyond the scope of our paper, which presents a first step to quantify the possible use of clock measurements in potential recovery. 





\paragraph{\textbf{Influence of the measurement noise.}}
\label{ssec:noiseeffect}

We have also investigated the effect of the noise levels applied to the synthetic data, see Tables~\ref{tab:effect_noise_Auv}--\ref{tab:effect_noise_Alps}, by using various standard deviations to simulate white noise of the measurements:~$\sigma_T = \{1, 0.1\}~\mcpsc$  for the clock measurements and~$\sigma_{\grdist} = \{1, 0.1, 0.01\}$~mGal for the gravimetric measurements. These results were obtained for the same conditions as in Section~\ref{sec:5}, i.e. 33 (resp. 32) clock data points and 4374 (resp. 4959) gravity data points for the Massif Central (resp. Alps).

%


We can see that adding clocks improves the potential recovery (smaller standard deviation ~$\sigma$ and bias~$\mu$ of the residuals) for both regions and whatever the noise of the gravimetric or clock measurements. 


\textcolor{black}{We observe that} decreasing the noise of the gravity data by up to two orders of magnitude only improves the standard deviation of the residuals~$\sigma$ of the recovered potential by comparatively small amounts (less than a factor 2). This is probably due to the fact that the covariance function does not reflect the gravity field correctly in these regions, combined with a limited data coverage. Note that the low frequency content in the covariance function (see above) is unlikely to be the main cause here, as the comparatively small reduction of~$\sigma$ is also observed when clocks are present in spite of the fact that they remove the low frequency trend (c.f. figures \ref{fig:Auv_EMC_Tdg_from_T33dg_56_15_4art} and \ref{fig:Alpes_EMC_Tdg_from_T32dg_50_3_4art}).

\begin{table*}[!htbp] 
	
	\sisetup{
		table-number-alignment = left,
		table-figures-integer = 0, 
		table-figures-decimal = 1,
		table-auto-round,
		table-sign-mantissa,
		table-sign-exponent,
		exponent-product =  \times, 
		output-exponent-marker 
	} 
	
	\captionsetup{width=1\linewidth}  
	\centering 
	
	\begin{footnotesize}
		\begin{tabular}{
				*{1}{C}
				*{8}{S[table-format = -1.1e2]}
			} 

			\toprule 
			
			\multicolumn{1}{c}{\diagbox{$\sigma_{T}$}{$\sigma_{\grdist}$}}
			& \multicolumn{2}{c}{5 mGal}
			& \multicolumn{2}{c}{1 mGal}
			& \multicolumn{2}{c}{0.1 mGal} 
			& \multicolumn{2}{c}{0.01 mGal} 
			\\
			\cmidrule(r){1-1}
			\cmidrule(r){2-3}
			\cmidrule(r){4-5}
			\cmidrule(r){6-7}
			\cmidrule(){8-9}
			
			{}
			& \multicolumn{1}{c}{$\mu$} & \multicolumn{1}{c}{$\sigma$}
			& \multicolumn{1}{c}{$\mu$} & \multicolumn{1}{c}{$\sigma$}
			& \multicolumn{1}{c}{$\mu$} & \multicolumn{1}{c}{$\sigma$}
			& \multicolumn{1}{c}{$\mu$} & \multicolumn{1}{c}{$\sigma$} 
			\\
			\cmidrule(r){1-1}
			\cmidrule(lr){2-2}\cmidrule(lr){3-3}
			\cmidrule(lr){4-4}\cmidrule(lr){5-5}
			\cmidrule(lr){6-6}\cmidrule(lr){7-7}
			\cmidrule(lr){8-8}\cmidrule(lr){9-9}
			
			\textrm{No clock}
			& 2.2E-1 & 3.7E-1
			& \color{red}4.1E-2 & \color{red}2.5E-1
			& 1.5E-1 & 1.7E-1
			& 2.6E-1 & 1.8E-1
			\\
			\cmidrule(lr){1-1}
			\cmidrule(lr){2-2}\cmidrule(lr){3-3}
			\cmidrule(lr){4-4}\cmidrule(lr){5-5}
			\cmidrule(lr){6-6}\cmidrule(lr){7-7}
			\cmidrule(lr){8-8}\cmidrule(lr){9-9}
			
			1~\mcpsc
			& -4.4E-3 & 2.8E-1
			& -1.8E-4 & 1.7E-1
			& -1.1E-2 & 1.6E-1
			& -2.0E-2 & 1.7E-1
			\\
			\cmidrule(lr){1-1}
			\cmidrule(lr){2-2}\cmidrule(lr){3-3}
			\cmidrule(lr){4-4}\cmidrule(lr){5-5}
			\cmidrule(lr){6-6}\cmidrule(lr){7-7}
			\cmidrule(lr){8-8}\cmidrule(lr){9-9}
			
			0.1~\mcpsc
			& -1.4E-2 & 2.0E-1
			& \color{red}-2.4E-3 & \color{red}7.3E-2
			& -6.7E-3 & 5.2E-2
			& -1.1E-3 & 4.8E-2
			\\
			
			\bottomrule 
		\end{tabular}
	\end{footnotesize}
	\caption{Noise level effect on the disturbing potential recovery in the Massif Central region. In red: results presented in Section~\ref{sec:5}. \textcolor{black}{Values are given in \mcpsc.}}
	\label{tab:effect_noise_Auv}
\end{table*}

\begin{table*}[!htbp] 
	
	\sisetup{
		table-number-alignment = left,
		table-figures-integer = 0, 
		table-figures-decimal = 1,
		table-auto-round,
		table-sign-mantissa,
		table-sign-exponent,
		exponent-product =  \times, 
		output-exponent-marker 
	} 
	
	\captionsetup{width=1\linewidth}  
	\centering 
	
	\begin{footnotesize}
		\begin{tabular}{
				*{1}{C}
				*{8}{S[table-format = -1.1e2]}
			} 

			\toprule 
			
			\multicolumn{1}{c}{\diagbox{$\sigma_{T}$}{$\sigma_{\grdist}$}}
			& \multicolumn{2}{c}{10 mGal}
			& \multicolumn{2}{c}{1 mGal}
			& \multicolumn{2}{c}{0.1 mGal} 
			& \multicolumn{2}{c}{0.01 mGal} 
			\\
			\cmidrule(r){1-1}
			\cmidrule(r){2-3}
			\cmidrule(r){4-5}
			\cmidrule(r){6-7}
			\cmidrule(){8-9}
			
			{}
			& \multicolumn{1}{c}{$\mu$} & \multicolumn{1}{c}{$\sigma$}
			& \multicolumn{1}{c}{$\mu$} & \multicolumn{1}{c}{$\sigma$}
			& \multicolumn{1}{c}{$\mu$} & \multicolumn{1}{c}{$\sigma$}
			& \multicolumn{1}{c}{$\mu$} & \multicolumn{1}{c}{$\sigma$} 
			\\
			\cmidrule(r){1-1}
			\cmidrule(lr){2-2}\cmidrule(lr){3-3}
			\cmidrule(lr){4-4}\cmidrule(lr){5-5}
			\cmidrule(lr){6-6}\cmidrule(lr){7-7}
			\cmidrule(lr){8-8}\cmidrule(lr){9-9}

			\textrm{No clock}
			& 5.8E-1 & 6.6E-1
			& \color{red}2.2E-1 & \color{red}3.9E-1
			& 2.1E-1 & 4.2E-1
			& 2.1E-1 & 4.2E-1
			\\
			\cmidrule(lr){1-1}
			\cmidrule(lr){2-2}\cmidrule(lr){3-3}
			\cmidrule(lr){4-4}\cmidrule(lr){5-5}
			\cmidrule(lr){6-6}\cmidrule(lr){7-7}
			\cmidrule(lr){8-8}\cmidrule(lr){9-9}
			
			1~\mcpsc
			& 1.8E-1 & 6.2E-1
			& 1.4E-1 & 3.4E-1
			& 1.2E-1 & 3.3E-1
			& 1.2E-1 & 3.3E-1
			\\
			\cmidrule(lr){1-1}
			\cmidrule(lr){2-2}\cmidrule(lr){3-3}
			\cmidrule(lr){4-4}\cmidrule(lr){5-5}
			\cmidrule(lr){6-6}\cmidrule(lr){7-7}
			\cmidrule(lr){8-8}\cmidrule(lr){9-9}
			
			0.1~\mcpsc
			& 2.0E-1 & 5.6E-1
			& \color{red}6.8E-2 & \color{red}1.7E-1
			& 4.7E-2 & 1.5E-1
			& 1.7E-2 & 1.6E-1
			\\
			
			\bottomrule 
		\end{tabular}
	\end{footnotesize}
	\caption{Noise level effect on the disturbing potential recovery in the Alps region. In red: results presented in Section~\ref{sec:5}. \textcolor{black}{Values are given in \mcpsc.}}
	\label{tab:effect_noise_Alps}
\end{table*}

When adding clocks the standard deviations are decreased by up to a factor 3.7 with low clock noise (0.1~\mcpsc or 1~cm) and a factor 1.5 with higher clock noise (1~\mcpsc or 10~cm). The effect is stronger in the Massif Central region than in the Alps. We attribute this again to the mismatch between the covariance function and the complex structure of the gravity field, which is larger in the Alps.
%

\textcolor{black}{Basically, the simulations put in evidence that the solutions depend on two types of errors, the measurement accuracy and the representation error. Indeed, if we increase the number of gravity data at high spatial resolution, we reduce the modelling error, which solves the problem of data interpolation; inversely, the modelling error will be more important if we have a poor coverage and gaps. But the quality of the covariance model is also reflected by the quality of the measurements as illustrated by the first column in Tables~\ref{tab:effect_noise_Auv}--\ref{tab:effect_noise_Alps} where we have used a high noise level for the gravity measurements, discussed in the next Section.}

Thus, optical clocks with just an accuracy of 1~\mcpsc (or 10~cm) are interesting no matter what the gravity data quality. With an accuracy of 0.1~\mcpsc (or 1~cm), we can expect a gain of up to a factor 4 in the estimated potential with respect to simulations using no clock data. \textcolor{black}{Of course, this gain depends on the number of clocks and the geometry of the clock coverage. For several tested configurations, we have remarked that it is possible to obtain the same gain in terms of rms with less \textcolor{black}{clocks} (e.g. about 10 clocks) but with a slightly larger bias. Additionally, different spatial distribution of the same number of clocks can degrade or improve the quality on the determination of $T$.}

\paragraph{\textbf{Aliasing of the very high-resolution components.}}
\label{ssec:aliasing}

\textcolor{black}{We have studied the aliasing of gravity variations at scales shorter than 10 km spatial resolution, that would be present in real data but under-sampled by the finite spatial density of the surveys. Errors in the topographic corrections may reach a few mGal for DTM (Digital terrain model) sampled at hundreds of meters resolution \citep{Tziavos_2009aa}, while local geological sources may lead to gravity signals up to $\sim~10$~mGal \citep{Yale_1998aa,Bondarescu_2012aa,Castaldo_2014aa}. Furthermore, we have analyzed the Bouguer gravity anomalies from the BGI database along profiles in the Massif Central and the Alps, and found, after smoothing the profiles at 10 km resolution, high-resolution components with rms $\sim$~1 mGal in the Massif Central, and $\sim$~3 mGal in the Alps. An order of magnitude of the corresponding geoid variations can be derived by assuming that the gravity signals at a given spatial scale are created by a point mass at the corresponding depth. We find that a 5~km width, 5~mGal (resp. 10~mGal) gravity anomaly corresponds to a 1.3~cm (resp. 2.6~cm) geoid variation, above the centimetric level indeed.}
	
\textcolor{black}{We simulate these previously neglected signals beyond 10~km resolution by increasing the noise level on the gravity data in our tests, up to 5 mGal in the Massif Central, and 10 mGal in the Alps. Note that these rms values are large with respect to the observed high-resolution variabilities in the data. 
As previously, numerical simulations are performed with and without adding clocks, and the results are presented in the first column of Table~\ref{tab:effect_noise_Auv}--\ref{tab:effect_noise_Alps}. We can see that decreasing the accuracy of the gravimetric measurements increases the residuals as compared to the previous solutions. This is due to the fact that the signal-to-noise ratio decreases, degrading the covariance function modelling. However, our previous conclusions on the benefit of clocks remain the same, even in the presence of significant signals at the shortest spatial scales.}


\section{Conclusions}
\label{sec:ccl}

Optical clocks provide a tool to measure directly the potential \textcolor{black}{differences} and determine the geopotential at high spatial resolution. We have shown that the recovery of the potential from gravity and clock data with the LSC method can improve the determination of geopotential at high spatial resolution, beyond what is available from satellites. Compared to a solution that does not use the clock data, the standard deviation \textcolor{black}{of the disturbing potential reconstruction} can be improved by a factor 3, and the bias can be reduced by up to 2 orders of magnitude with only a few tens of clock data. This demonstrates the benefit of this new potential geodetic observable, which could be put in practice in the medium term when the first transportable optical clocks and appropriate time transfer methods will be developed \citep[see][]{Bongs_2015aa,Lisdat_2016aa,Deschenes_2016aa,Vogt_2016aa}. Since clocks are sensitive to low frequencies of the gravity field, this method is particularly well-adapted in hilly and mountainous regions for which the gravity coverage is more sparsely distributed, allowing to fill areas not covered by the classical geodetic observables (gravimetric measurements). Additionally, adding new observables helps to reduce the modelling errors, e.g. coming from a mismatch between the covariance function used and the real gravity field.

\textcolor{black}{In the same way, GPS and leveling data have been used, in combination with gravity data, to derive high-resolution gravimetric geoids \citep{Kotsakis_1999aa, Duquenne_1999aa, Denker_2000aa, Duquenne_2005aa, Nahavandchi_2006aa}. Using clocks is, however, different from performing GPS and leveling measurements. They provide an information of similar nature as the gravity data, in contrast with these geometric observations. The latter are affected by different sources of errors \citep[e.g.][]{Duquenne_1998aa, Marti_2001aa}, and quite expensive in the case of leveling campaigns. We can expect that clocks could help identify and reduce errors in the gravity and GPS/leveling  through their joint analysis for geopotential determination.
Beyond the application considered in this work, the clocks can also contribute to the unification of height systems realizations \citep{Shen_2011aa, Denker_2013aa, Shen_2016aa, Kopeikin_2016aa, Takano_2016aa}, connecting distant points to a high-resolution reference potential network.}

To our knowledge, this is the first detailed quantitative study of the improvement in field determination that can be expected from chronometric geodesy observables. It provides first estimates and paves the way for future more detailed and in depth works in this promising new field.

To overcome some limitations in the a priori model, as discussed in the previous section, we intend in a forthcoming work to investigate in more details the imperfections of the covariance function model. Moreover, as the gravity field is in reality non-stationary in mountainous areas or near the coast, some numerical tests with non-stationary covariance functions will be conducted. Another promising source of improvement could be the optimization of the positioning of the clock data. For example, the correlation lengths and the variations of the gravity field could be used as constraints. A genetic algorithm could also be considered to solve this location problem. Finally, it will be interesting to focus on the improvement of the potential recovery quality by combining other types of observables such as leveling data and gradiometric measurements. As knowledge of the geopotential provides access to height differences, this could be a way to estimate errors of the GNSS technique for the vertical positioning, or contribute to regional height systems unification.

\section*{Acknowledgements}

We thank René Forsberg for providing us the fortran code of the logarithmic covariance function.
We gratefully acknowledge financial support from Labex FIRST-TF, ERC AdOC (grant n\si{\degree}~617553 and EMRP ITOC (EMRP is jointly funded by the EMRP participating countries within EURAMET and the European Union).
We thank Olivier Jamet and Matthias Holschneider for discussions about the collocation method. 
\textcolor{black}{We thank Gwendoline Pajot-Métivier for discussions on high-resolution gravity signals. 
We thank anonymous reviewers and the associate editor for their useful comments on this manuscript.}

\renewcommand*\bibfont{\small}

\setlength\bibsep{0pt} 

\bibliographystyle{spbasic}      


\begin{appendices}{\normalsize}

\section{Covariance function}
\label{anx:defFCF}

Let us consider two points~$P$ and~$Q$ with the Cartesian coordinates~$(x_P,y_P,z_P)$ and~$(x_Q,y_Q,z_Q)$, respectively. To compute the ACF and CCF of the disturbing potential~$T$ and its derivatives,  \citet{Forsberg_1987aa} proposed a planar attenuated logarithm covariance model with upward continuation that can be expressed in the generic form
\begin{align}
& C(x,y,z_1+z_2) = S \sum_{i=0}^{3} \lambda_i K(x,y,z_i)
\end{align}
with%
\begin{subequations}
\begin{align}
& x = x_Q-x_P \;, \quad  y = y_Q-y_P \\
& z_i = z_P + z_Q + \alpha_i \\
& \alpha_i = \alpha + i\,\beta \\
& \lambda_i = \{1, -3, 3, -1 \} \\
& S = C_0 \log^{-1}\fracp*{\alpha_1^3 \, \alpha_3}{\alpha_0 \alpha_2^3}
\end{align}
\end{subequations}
This model is characterized by three parameters:~$C_0$ the variance of the gravity disturbance~$\grdist$ and two scale factors acting as high and low frequency attenuators:~$\alpha$ the shallow depth parameter and~$\beta$ the compensating depth, respectively. 
The function~$K=K(x,y,z_i)$ is logarithmic function modelling the covariances between the gravity field quantities.
For example, by putting~$r_i = \sqrt{d^2 + \alpha_i^2}$ and~$d = \sqrt{x^2 + y^2}$, the ACF of~$\grdist$ and~$T$ can be evaluated respectively with 
\begin{align}
& K = -\log (\alpha_i + r_i) \\
& K = \frac{3}{4} z_i r_i + \left( \frac{r_i^2}{4} - \frac{3}{4} z_i^2 \right) \log (z_i + r_i)
\end{align}
and the CCF between~$T$ and~$\grdist$ with
\begin{align}
K  & = r_i - z_i \log (z_i + r_i)
\end{align}
%

\end{appendices}

\end{document}